\documentclass[a4paper,11pt]{article}

\usepackage{color}
\usepackage{ulem}
\usepackage{hyperref}
\usepackage{enumerate}
\usepackage{amsmath}
\usepackage{graphicx}
\usepackage{float}
\usepackage{subfigure}

\def \be {\begin{equation}}
\def \ee {\end{equation}}
\def \ba {\begin{array}}
\def \ea {\end{array}}
\def \bea {\begin{eqnarray}}
\def \eea {\end{eqnarray}}

\def \ble {\begin{widetext}\begin{equation}}
\def \ele {\end{equation}\end{widetext}}
\def \blea {\begin{widetext}\begin{eqnarray}}
\def \elea {\end{eqnarray}\end{widetext}}

\def \nn {\nonumber}

\def \blea {\begin{widetext}\begin{eqnarray}}
\def \elea {\end{eqnarray}\end{widetext}}

\def \p {\partial}
\def \mP {\mathcal{P}} 

\begin{document}

\title{Imaginary part of timelike entanglement entropy}

\author{Wu-zhong Guo\footnote{wuzhong@hust.edu.cn}~, Jin Xu\footnote{d202180103@hust.edu.cn}}

\date{}
\maketitle

\vspace{-10mm}
\begin{center}
{\it School of Physics, Huazhong University of Science and Technology,\\
 Wuhan, Hubei
430074, China
\vspace{1mm}
}
\vspace{10mm}
\end{center}

\begin{abstract}
In this paper, we explore the imaginary part of the timelike entanglement entropy. In the context of field theory, it is more appropriate to obtain the timelike entanglement entropy through the Wick rotation of the twist operators. It is found that, in certain special cases, the imaginary part of the timelike entanglement entropy is related to the commutator of the twist operator and its first-order temporal derivative. To evaluate these commutators, we employ the operator product expansion of the twist operators, revealing that the commutator is generally \textit{universal} across most scenarios. However, in more general cases, the imaginary part of the timelike entanglement entropy proves to be more complex. We compute the commutator of the twist operators along with its higher-order temporal derivatives. Utilizing these results, we derive a modified formula for the imaginary part of the timelike entanglement entropy. Furthermore, we extend this formula to the case of strip subregion in higher dimensions. Our analysis shows that for the strip geometry, the imaginary part of the timelike entanglement entropy is solely related to the commutators of the twist operator and its first-order temporal derivative. The findings presented in this paper provide valuable insights into the imaginary part of timelike entanglement entropy and its physical significance.
\end{abstract}
\tableofcontents

\maketitle

\section{Introduction}
The study of quantum field theory (QFT) reveals that entanglement is a crucial tool for probing its fundamental structure. We can employ a variety of measures to quantify entanglement, with entanglement entropy (EE) being one of the most thoroughly investigated. Both analytical and numerical calculations have significantly enhanced our comprehension of EE, as referenced in various studies \cite{Holzhey:1994we,Srednicki:1993im,Vidal:2002rm,Calabrese:2004eu,Calabrese:2009qy,Casini:2009sr}. Intriguingly, within the AdS/CFT framework \cite{Maldacena:1997re,Gubser:1998bc,Witten:1998qj}, entanglement entropy has been discovered to correlate with minimal surfaces in the corresponding spacetime \cite{Ryu:2006bv,Hubeny:2007xt}, adhering to a principle akin to the area law observed in black holes.

 The state of QFT is characterized by the density matrix $\rho$. By partitioning the system into subsystem $A$ and its complementary $\bar A$, one can introduce the reduced density matrix $\rho_A := \text{tr}_{\bar A} \rho$. EE can then be defined as the Von Neumann entropy $S(\rho_A):=-\text{tr} \rho_A\log \rho_A$. The $n$-th R\'enyi entropy $S^{(n)}(\rho_A) := \frac{1}{1-n} \log \text{tr} \rho_A^n$ can be evaluated by path integral on  $n$-copied theory glued together along the subsystem $A$. By analytically continuing $n$ to complex numbers, the EE can be evaluated by the limit $S(\rho_A) = \lim_{n \to 1} S^{(n)}(\rho_A)$.

 In QFTs, it is more appropriate to understand the Rényi entropy through the use of twist operators. Specifically, the expression  $tr(\rho_A)^n$ can be taken as correlator involving  twist operators for the $n$-copied CFT$_n$. If taking  $A$ to be an interval in 2-dimensional QFTs, we have 
\bea\label{twist}
tr\rho^n_A=\langle \Psi|\sigma_n(\tau,x)\tilde{\sigma}_n(\tau',x')|\Psi\rangle,
\eea
where $(\tau,x)$ and $(\tau',x')$ are coordinates of the endpoints of $A$,  $|\Psi\rangle:=|\psi\rangle_1\otimes ...|\psi\rangle_i \otimes ... |\psi\rangle_n$, the subscripts $i$ label the $i$-th copy. In 2-dimensional CFTs the twist operators can be taken as local primary operators with the conformal dimension $h_n=\bar h_n=\frac{c}{24}(n-\frac{1}{n})$\cite{Calabrese:2004eu,Calabrese:2009qy}.

In past research on EE, the primary focus has been on the scenario where $A$ is a spacelike separation. Recently, there have been efforts to extend this focus to timelike scenarios \cite{Doi:2022,Doi:2023}. The approach involves interchanging the roles of time and space coordinates. Consequently, timelike entanglement entropy is expected to be complex-valued. From the perspective of twist operators, it is more appropriate to understand timelike EE through the standard Wick rotation in QFTs, given by $\tau\to it+\epsilon$\cite{Guo:2024lrr}. Since in general cases, the timelike EE is complex-valued, and it may be interpreted as pseudo entropy\cite{Nakata:2020,Murciano:2020}. For more recent developments, one could refer to \cite{Li:2023yyl,Das:2023yyl,Chu:2023zah,Jiang:2023loq,Chen:2023gnh,He:2023ubi,Narayan:2023zen,Carignano:2023,Basu:2024,Carignano:2024,Anegawa:2024,Heller:2024,Bou-Comas:2024}.

Furthermore, it has been observed in \cite{Guo:2024lrr} that in many instances within 2-dimensional conformal field theories (CFTs), the timelike EE is uniquely determined by a linear combination of the spacelike EE and its first-order temporal derivative, 
\begin{align}\label{S_separation}
&S(t,x;t,x')\nn\\
=&\frac{1}{4}(S(0,-u;0,-u')+S(0,-u;0,v')+S(0,v;0,-u')+S(0,v;0,v'))\nn\\
&+\frac{1}{4}\int_{-u'}^{v'}\text{d}y \p_{t'}S(0,-u;0,y)+\frac{1}{4}\int_{-u'}^{v'}\text{d}y \p_{t'}S(0,v;0,y)\nn\\
&+\frac{1}{4}\int_{-u}^{v}\text{d}y \p_{t}S(0,y;0,-u')+\frac{1}{4}\int_{-u}^{v}\text{d}y \p_{t}S(0,y;0,v')\nn\\
&+\frac{1}{4}\int_{-u}^{v}\text{d}y \int_{-u'}^{v'}\text{d}y'\p_{t}\p_{t'}S(0,y;0,y'),
\end{align}
where $u=t-x$, $v=t+x$, $u'=t'-x'$ and $v'=t'+x'$. $S(t,x;t,x')$ denotes the EE for the interval  between the points $(t,x)$ and $(t',x')$.  And the time derivative term should be understood as taking the derivative first and then taking $t\to 0$ and $t'\to 0$.

One of the most intriguing aspects of timelike EE is its complex-valued nature. It is crucial to understand the imaginary part of the timelike EE, The relation given by Eq.(\ref{S_separation}) provides a means to do this. The imaginary part of the timelike entanglement entropy is entirely derived from the middle four terms, particularly those involving  $\p_{t}S(0,x;0,x')$ or $\p_{t'}S(0,x;0,x')$. Due to the relationship between R\'enyi entropy and twist operators, we have further uncovered a connection between their imaginary parts and the first-order time derivative of the commutator of twist operators $\langle[\dot{\sigma}_n(0,x),\tilde{\sigma}_n(0,x')]\rangle$. Note that since the right hand side of (\ref{S_separation}) are the results on a certain Cauchy surface, the commutator is on  a certain Cauchy surface $t=0$.

In this paper, we first provide a concise extension of the research presented in \cite{Guo:2024lrr}, elucidating the influence of various analytic continuation methods on the computation of timelike EE.  We will explain why the choice of the $i\epsilon$ prescription is ultimately adopted. Furthermore, we will extend the result of (\ref{S_separation}) in \cite{Guo:2024lrr}, which was originally derived for a specific Cauchy surface at $t=0$, to an arbitrary Cauchy surface at $t=t_0$. And we find that regardless of the choice of the Cauchy surface, the imaginary part always arises from the middle four terms of (\ref{S_separation}).

We further study the commutator $\langle[\dot{\sigma}_n(t_0,x),\tilde{\sigma}_n(t_0,x')]\rangle_{\Psi}$. For the vacuum state, thermal state, and states dual to pure AdS$_3$, one can directly compute the commutator using the results of EE. These results align with previous analyses \cite{Guo:2024lrr}. This consistency is anticipated, as the examples mentioned all satisfy equation (\ref{S_separation}). Additionally, one can evaluate the commutators in CFTs for general states by utilizing the operator product expansion (OPE) of twist operators. We find that the commutator $\langle[\dot{\sigma}_n(t_0,x),\tilde{\sigma}_n(t_0,x')]\rangle_{\Psi}$ is universal, that is independent with the state $|\Psi\rangle$. This universality can account for the seemingly consistent imaginary part of the timelike EE across many states. Specifically, for a general state, we have
\begin{align}
&\frac{\langle \Psi|[\dot{\sigma}_n(t_0,x),\tilde{\sigma}_n(t_0,x')]|\Psi\rangle}{\langle \Psi| \sigma_n(t_0,x)\tilde{\sigma}_n(t_0,x')|\Psi\rangle}=-8h_n\pi i\delta(x-x').
\end{align}
On the other hand, for general states in 2D CFT, the first derivative of its commutator with respect to the spatial coordinate x is
\begin{align}
&\frac{\langle \Psi|[\p_x\sigma_n(t_0,x),\tilde{\sigma}_n(t_0,x')]|\Psi\rangle}{\langle \Psi| \sigma_n(t_0,x)\tilde{\sigma}_n(t_0,x')|\Psi\rangle}=0.
\end{align}
Therefore, the partial derivative with respect to the spatial coordinate $x$ does not introduce a new imaginary part.

However, we must emphasize that the imaginary part of the timelike entanglement entropy should not be constant for most general states. These new imaginary components are expected to be related to the commutators involving higher-order temporal derivatives. These commutators can be calculated in 2D CFTs, as demonstrated in Section \ref{section_general_results}. In the most general cases, the relation between timelike and spacelike EE expressed in (\ref{S_separation}) should also be modified. The right-hand side should incorporate higher-order temporal derivatives. Additionally, we propose a new formula that connects the imaginary part of the timelike EE to the spacelike EE and its temporal derivatives.

Furthermore, we have also explored the discussion of high-dimensional examples. For the timelike strip subregion, one can evaluate the timelike entanglement entropy using the holographic entanglement entropy formula or by analytically continuing the spacelike result. Additionally, we can calculate the commutator of the twist operator and its first-order temporal derivative. Utilizing this result, we find that the imaginary part of the timelike entanglement entropy for the strip can be associated with these commutators, providing a new perspective on understanding the imaginary part of timelike entanglement entropy in higher-dimensional examples.

Finally, we discuss the timelike entanglement entropy and the commutator of the twist operator in the context of entanglement spectra \cite{Li:2008kda,Calabrese2008,Hung:2011nu,Guo:2020rwj,Guo:2020roc}. Such discussions are feasible through the parameter dependence of the entanglement spectra \cite{Guo:2023p}, as we can define $\mP_{t}(\lambda) := \sum_i \frac{\partial \lambda_i}{\partial t} \delta(\lambda - \lambda_i)$. Since the entanglement spectra and the R\'enyi entropy are Laplace transforms of one another, their results are naturally consistent.

he remainder of the paper is organized as follows. In Section \ref{section_2}, we analyze the effects of different analytic continuation methods and explain why the $i\epsilon$ prescription should be adopted. We also investigate the origin of the imaginary part of the timelike entanglement entropy across various Cauchy surfaces and its relationship to the commutators of the twist operator and its first-order temporal derivative. Following this, in Section \ref{section_example}, we compute the commutators for the vacuum state, thermal state, and states dual to pure AdS$_3$. In Section \ref{section_General}, utilizing the OPE of twist operators, we discuss the imaginary part of the timelike EE for general states and find that it originates solely from the identity operator and operators with fractional conformal dimensions. We propose a modified formula for the imaginary part of the timelike EE, expressing it in terms of the spacelike EE and its derivatives on a specific Cauchy surface at $t=t_0$
In Section \ref{section_Higher dimension}, we explore the strip in higher dimensions and evaluate the commutator involving the first-order derivative of the twist operator, constructing the formula for the imaginary part of the timelike EE in this context. In Section \ref{section_spectra}, we present consistent results concerning the spectra. Finally, Section \ref{section_Conclusion} offers concluding remarks, with detailed calculations provided in the appendices.

\section{Timelike and spacelike entanglement entropy relation}\label{section_2}

\subsection{Timelike entanglement entropy via analytical continuation}
EE can be evaluated through the correlators of twist operators in Euclidean QFTs. The correlators in Lorentzian QFTs can be obtained via a Wick rotation, $\tau \rightarrow it + \epsilon$, from the Euclidean QFT results. This provides a practical method to define timelike EE \cite{Guo:2024lrr}. Different analytic continuations lead to distinct operator orderings in the correlator \cite{Hartman:2015Causality}. In the following, we will primarily focus on two-point correlation functions with timelike separation\footnote{In the replica method for evaluating EE in QFTs, we typically work within the framework of Euclidean QFT, where the twist operators are defined on a constant Euclidean time slice. In Appendix \ref{app1}, we briefly discuss the significance of the correlators of twist operators involving time derivatives in both Euclidean and Lorentzian QFTs.}.

For simplicity and to aid intuitive understanding, we will specifically demonstrate the results for the EE of a single interval in the vacuum state. In the case of the vacuum state $|\Psi\rangle = |0\rangle$ in 2D conformal field theories (CFTs), the R\'enyi entropy is given by
\begin{align}\label{Sn}
S_n(\tau,x;\tau',x')&=\frac{1}{1-n}\log\langle \sigma_n(\tau,x)\tilde{\sigma}_n(\tau',x')\rangle\nn
\end{align}
With the Wick rotation  we have
\begin{align}
S_n(t,x;t',x')&=\frac{2h_n}{n-1}\log{[(\tau-\tau')^2+(x-x')^2]}|_{\tau\to it+\epsilon,\tau'\to it'+\epsilon'}\nn \\
\phantom{S_n(t,x;t',x')}&=\frac{2h_n}{n-1}\log{[\Delta s^2+2i(t-t')(\epsilon-\epsilon')]},
\end{align}
where $\Delta s^2=-(t-t')^2+(x-x')^2$ and $\epsilon,\epsilon'>0$. 

Now We will have the following four different cases.

In Case 1, when $t<t'$, $\epsilon<\epsilon'$, and when $t>t'$, $\epsilon>\epsilon'$, the correlator would be fully time-ordered, then we have
\begin{align}
S_n(t,x;t',x')=S_n(t',x';t,x)=\frac{c}{12}(1+\frac{1}{n})\log{[-\Delta s^2]}+\frac{i\pi c}{12}(1+\frac{1}{n}),
\end{align}
which means that the timelike R\'enyi entropy is independent of the time ordering of the two points.

In Case 2, when $t<t'$, $\epsilon>\epsilon'$, or when $t>t'$, $\epsilon<\epsilon'$, the correlator is fully anti-time-ordered,  then we have
\begin{align}
S_n(t,x;t',x')=S_n(t',x';t,x)=\frac{c}{12}(1+\frac{1}{n})\log{[-\Delta s^2]}-\frac{i\pi c}{12}(1+\frac{1}{n}).
\end{align}
In this case, the timelike R\'enyi entropy is also independent of the time ordering of the two points.

In Case 3, when $t<t'$, $\epsilon>\epsilon'$, or when $t>t'$, $\epsilon>\epsilon'$,  we have
\begin{align}
S_n(t,x;t',x')=S_n(t',x';t,x)^*=\frac{c}{12}(1+\frac{1}{n})\log{[-\Delta s^2]}+\frac{i\pi c}{12}(1+\frac{1}{n}).
\end{align}
In this case, the result is dependent with the orderings of the two points. This is the case that is studied in \cite{Guo:2024lrr}.  

In Case 4, when $t<t'$, $\epsilon<\epsilon'$, or when $t>t'$, $\epsilon<\epsilon'$, we have
\begin{align}
S_n(t,x;t',x')=S_n(t',x';t,x)^*=\frac{c}{12}(1+\frac{1}{n})\log{[-\Delta s^2]}-\frac{i\pi c}{12}(1+\frac{1}{n}).
\end{align}

As seen from the above results, in Case 1 and Case 2, the ordering of the two points with timelike separation is unimportant. However, in Case 3 and Case 4, the ordering becomes significant for the imaginary terms.

The imaginary part of the timelike Rényi entropy can be obtained by Im$[S_n(t,x;t',x')] = \frac{1}{2}[S_n(t,x;t',x') - S_n(t,x;t',x')^*]$. As we will show, this imaginary term is closely related to the local commutator of the twist operators and its temporal derivative. One could define the timelike Rényi entropy using any of the four cases mentioned above. However, the relation between timelike and spacelike EE  (\ref{S_separation}) is derived from Case 3 \cite{Guo:2024lrr}. In the following discussion, we will adopt Case 3 as the definition of timelike R\'enyi entropy, though similar relations between timelike and spacelike EE can also be obtained using other definitions. Since in Case 3, $\epsilon > \epsilon'$ always holds, we can redefine $\epsilon - \epsilon'$ as an infinitesimal positive quantity $\epsilon$.


\subsection{The imaginary term for different separations}
Previous research has shown that the two-point correlators $\langle\phi(t,x)\phi(t',x')\rangle$ have a form similar to the R\'enyi entropy of a single interval in the vacuum state, where $\phi(t,x)$ is a massless free scalar field. Using this similarity, one can demonstrate that the timelike EE can be expressed as a linear combination of the spacelike entanglement entropy and its first-order temporal derivative on the Cauchy surface $t = 0$ \cite{Guo:2024lrr}. This relation can be generalized to the states with holographic dual.

We can generalize the relation (\ref{S_separation}) to an arbitrary Cauchy surface at $t = t_0$. It can also be shown that this relation holds not only for two points with timelike separation but also for those with spacelike separation. Therefore, this provides a unified framework for understanding entanglement for arbitrary intervals in Minkowski spacetime.

The EE of a single interval in vacuum state  with any separation of two points $(t,x)$ and $(t',x')$  can be expressed by spacelike EE and their first-order time derivative on any Cauchy surface $t = t_0$,
\begin{align}\label{Sn_separation}
&S(t,x;t',x')\nn\\
=&\frac{1}{4}[S(t_0,-u;t_0,-u')+S(t_0,-u;t_0,v')+S(t_0,v;t_0,-u')+S(t_0,v;t_0,v')]\nn\\
&+\frac{1}{4}\int_{-u'}^{v'}\text{d}y \p_{t'}S(t_0,-u;t_0,y)+\frac{1}{4}\int_{-u'}^{v'}\text{d}y \p_{t'}S(t_0,v;t_0,y)\nn\\
&+\frac{1}{4}\int_{-u}^{v}\text{d}y \p_{t}S(t_0,y;t_0,-u')+\frac{1}{4}\int_{-u}^{v}\text{d}y \p_{t}S(t_0,y;t_0,v')\nn\\
&+\frac{1}{4}\int_{-u}^{v}\text{d}y \int_{-u'}^{v'}\text{d}y'\p_{t}\p_{t'}S(t_0,y;t_0,y'),
\end{align}
where $u=(t-t_0)-x$, $v=(t-t_0)+x$, $u'=(t'-t_0)-x'$ and $v'=(t'-t_0)+x'$.  And the time derivative term should be understood as taking the derivative first and then taking $t\to t_0$ and $t'\to t_0$. For example, $\p_{t}\p_{t'}S(0,x;0,x')$ should be understood as $\{\p_{t}\p_{t'}S(t,x;t',x')\}|_{t\to t_0,t'\to t_0}$. 

If the two points have a timelike separation, an imaginary term will appear. In contrast, for spacelike separation, there will be no imaginary term. Our focus is on how the imaginary term arises in the timelike case. In fact, these imaginary terms originate from the middle four terms in (\ref{Sn_separation}).

\begin{figure}
\centering
\subfigure[]{\includegraphics[scale=0.7]{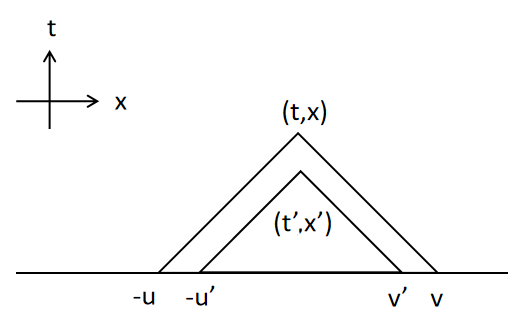}}
\subfigure[]{\includegraphics[scale=0.7]{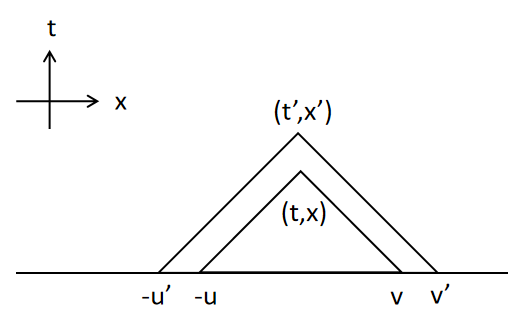}}
\subfigure[]{\includegraphics[scale=0.7]{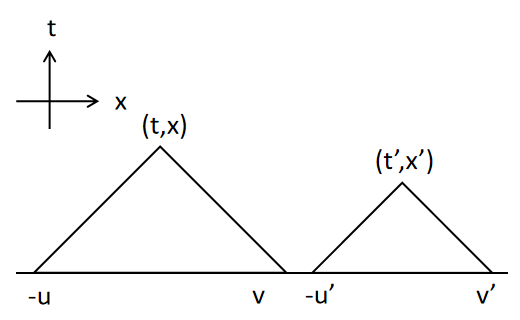}}
\subfigure[]{\includegraphics[scale=0.7]{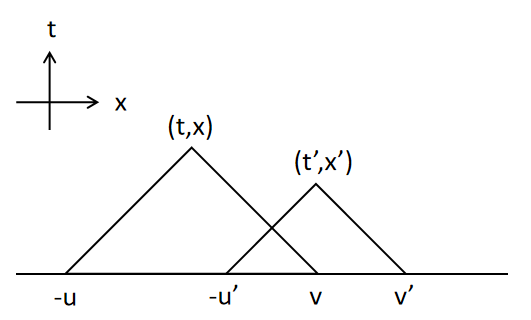}}
\caption{(a) gives the timelike separation $t>t'$. (b) gives the timelike separation $t<t'$. (c) gives the spacelike separation, where the light cones of its two points have no intersection on the Cauchy surface $t=t_0$. (d) gives the spacelike separation, where the light cones of its two points intersect on the Cauchy surface $t=t_0$.}
\label{separation}
\end{figure}

Let us consider the following four different cases shown in Fig.\ref{separation}.
For timelike separation $t>t'$ shown in Fig.\ref{separation} (a), the imaginary term comes from
\begin{align}
\frac{1}{4}\int_{-u}^{v}\text{d}y \p_{t}S(t_0,y;t_0,-u')+\frac{1}{4}\int_{-u}^{v}\text{d}y \p_{t}S(t_0,y;t_0,v')=\frac{i\pi c}{6}.
\end{align}

For timelike separation $t<t'$ shown in Fig.\ref{separation} (b), the imaginary term comes from
\begin{align}
\frac{1}{4}\int_{-u'}^{v'}\text{d}y \p_{t'}S(t_0,-u;t_0,y)+\frac{1}{4}\int_{-u'}^{v'}\text{d}y \p_{t'}S(t_0,v;t_0,y)=-\frac{i\pi c}{6}.
\end{align}

For spacelike separation shown in Fig.\ref{separation} (c), there is no imaginary term,
\begin{align}
&\frac{1}{4}\int_{-u'}^{v'}\text{d}y \p_{t'}S(t_0,-u;t_0,y)=\frac{1}{4}\int_{-u'}^{v'}\text{d}y \p_{t'}S(t_0,v;t_0,y)\nn\\
&=\frac{1}{4}\int_{-u}^{v}\text{d}y \p_{t}S(t_0,y;t_0,-u')=\frac{1}{4}\int_{-u}^{v}\text{d}y \p_{t}S(t_0,y;t_0,v')=0.
\end{align}

For spacelike separation in Fig.\ref{separation} (d), the imaginary terms cancel each other out,
\begin{align}
&\frac{1}{4}\int_{-u'}^{v'}\text{d}y \p_{t'}S(t_0,-u;t_0,y)+\frac{1}{4}\int_{-u'}^{v'}\text{d}y \p_{t'}S(t_0,v;t_0,y)\nn\\
&+\frac{1}{4}\int_{-u}^{v}\text{d}y \p_{t}S(t_0,y;t_0,-u')+\frac{1}{4}\int_{-u}^{v}\text{d}y \p_{t}S(t_0,y;t_0,v')=0.
\end{align}
The detailed calculations are provided in Appendix \ref{Details_separations}.

As expected, the imaginary term vanishes for spacelike separations. Specifically, since (c) and (d) represent the results of using different Cauchy surfaces for the same spacelike separation of $(t,x)$ and $(t',x')$, and these results are consistent, this demonstrates that our generalization to any Cauchy surface (\ref{Sn_separation}) is self-consistent.

In both cases (a) and (b), we observe that the imaginary component arises from the first-order time derivative of the EE. In the following, we will demonstrate that this imaginary term is closely connected to the commutator of twist operators and its time derivative.

\section{Commutators of twist operators}\label{section_example}

We will further investigate the first-order temporal derivative terms of EE. By the definition of R\'enyi entropy, we have
\begin{align}
\p_t S_n(t,x;t',x')=\frac{1}{1-n}\frac{\langle \Psi|\dot{\sigma}_n(t,x)\tilde{\sigma}_n(t',x')|\Psi\rangle}{\langle \Psi|\sigma_n(t,x)\tilde{\sigma}_n(t',x')|\Psi\rangle}.
\end{align}
Therefore, the relationship between the twist operators and the first-order temporal derivative of EE is
\begin{align}
\p_t S(t,x;t',x')=\lim_{n\to1}\frac{1}{1-n}\frac{\langle \Psi|\dot{\sigma}_n(t,x)\tilde{\sigma}_n(t',x')|\Psi\rangle}{\langle \Psi|\sigma_n(t,x)\tilde{\sigma}_n(t',x')|\Psi\rangle}.
\end{align}
Based on the formula (\ref{Sn_separation}) discussed in the previous section, the imaginary part of timelike EE entropy is
\begin{align}\label{Im_0}
2 \text{Im} S(t,x;t',x')&= \frac{1}{4}\int_{-u}^{v}\text{d}y\lim_{n\to1}\frac{1}{1-n} \frac{\langle \Psi|[\dot{\sigma}_n(t_0,y),\tilde{\sigma}_n(t_0,-u')]|\Psi\rangle}{\langle\Psi|\sigma_n(t_0,y)\tilde{\sigma}_n(t_0,-u')|\Psi\rangle}\nn\\
&+\frac{1}{4}\int_{-u}^{v}\text{d}y\lim_{n\to1}\frac{1}{1-n} \frac{\langle \Psi|[\dot{\sigma}_n(t_0,y),\tilde{\sigma}_n(t_0,v')]|\Psi\rangle}{\langle \Psi|\sigma_n(t_0,y)\tilde{\sigma}_n(t_0,v')|\Psi\rangle}\nn\\
&+\frac{1}{4}\int_{-u'}^{v'}\text{d}y\lim_{n\to1}\frac{1}{1-n} \frac{\langle \Psi|[\dot{\sigma}_n(t_0,-u),\tilde{\sigma}_n(t_0,y)]|\Psi\rangle}{\langle \Psi|\sigma_n(t_0,-u)\tilde{\sigma}_n(t_0,y)|\Psi\rangle}\nn\\
&+\frac{1}{4}\int_{-u'}^{v'}\text{d}y\lim_{n\to1}\frac{1}{1-n} \frac{\langle \Psi|[\dot{\sigma}_n(t_0,v),\tilde{\sigma}_n(t_0,y)]|\Psi\rangle}{\langle \Psi|\sigma_n(t_0,v)\tilde{\sigma}_n(t_0,y)|\Psi\rangle}.
\end{align}
The above statement holds true for any separation that satisfies equation (\ref{Sn_separation}). For spacelike separation, the result is 0. For timelike separation of two points $(t,x)$ and $(t',x')$, the first two terms or the last two terms in the above equation will be 0 depending on whether $t>t'$ or $t'>t$ as we discussed in the previous section.

For the vacuum state, thermal state, and states dual to pure $\text{AdS}_3$, the relation (\ref{Sn_separation}) holds, with $\text{Im} S(t,x;t',x') = \frac{i\pi c}{6}$. To obtain this result for the imaginary part, we expect that the commutator between the twist operator and its first-order temporal derivative satisfies the relation \cite{Guo:2024lrr}
\begin{align}\label{commutator_0}
\frac{\langle \Psi|[\dot{\sigma}_n(t_0,x),\tilde{\sigma}_n(t_0,x')]|\Psi\rangle}{\langle \Psi|\sigma_n(t_0,x)\tilde{\sigma}_n(t_0,x')|\Psi\rangle}=\frac{2c}{3}(1-n)i\pi\delta(x-x')+O(1-n)^2.
\end{align}
The commutator should be valid for the three states mentioned above. In the following section, we will directly calculate the commutator involving the twist operator and its temporal and spatial derivatives.

\subsection{The commutator of twist operators in vacuum state}\label{com_twist}
We would like to firstly study the commutator of twist operators without temporal and spatial derivatives.
The two-point correlator of twist operators in the vacuum state is universal, 
\bea\label{Vacuum state_0}
 \langle \sigma_n(t,x)\tilde{\sigma}_n(t',x')\rangle=[-(u-u'-i\epsilon)]^{-2h_n}[(v-v'-i \epsilon)]^{-2\bar h_n}.
\eea
If the separation between $(t,x)$ and $(t',x')$ is spacelike, it can be shown 
\begin{align}
\langle [\sigma_n(t,x),\tilde{\sigma}_n(t',x')]\rangle=0.
\end{align}
One can also verify this directly using the result (\ref{Vacuum state_0}). This is required by microcausality QFTs. However, it is possible that the commutator may yield a delta function when $(t,x)\to (t',x')$.

One could also consider the commutator for timelike separation. Here, we focus only on the result for $n \sim 1$. By performing a Taylor expansion of (\ref{Vacuum state_0}) in terms of $(n-1)$, the result is
\begin{align}\label{Vacuum state_0_1}
\langle \sigma_n(t,x)\tilde{\sigma}_n(t',x')\rangle=1-2h_n\log{[-(u-u'-i\epsilon)(v-v'-i \epsilon)]}+O((n-1))^2).
\end{align}
For spacelike separation, (\ref{Vacuum state_0_1}) becomes
\begin{align}
\langle \sigma_n(t,x)\tilde{\sigma}_n(t',x')\rangle=1-2h_n\log{\Delta s^2}+O((n-1)^2),
\end{align}
where $\Delta s^2=-(t-t')^2+(x-x')^2$. For spacelike separation, as $\log\Delta s^2$ is real, the $k$-th order term of $(n-1)$ is propotional to $ (n-1)^k(\log\Delta s^2)^k$, which is also real. Thus, we have
\begin{align}
\langle [\sigma_n(t,x),\tilde{\sigma}_n(t',x')]\rangle=0.
\end{align}
This holds true for expanding $(n-1)$ to arbitrary order as long as it is for spacelike separation. The commutator is vanishing consistent with microcausality.

If the separation is timelike and $t>t'$, (\ref{Vacuum state_0_1}) becomes
\begin{align}
\langle \sigma_n(t,x)\tilde{\sigma}_n(t',x')\rangle=1-2h_n(\log(-\Delta s^2)+i\pi)+O((n-1)^2).
\end{align}
The commutator in this case is
\begin{align}\label{Vacuum state_0_timelike}
\langle [\sigma_n(t,x),\tilde{\sigma}_n(t',x')]\rangle=-4h_n \pi i+O((n-1)^2).
\end{align}
In this paper, we mainly focus on the case where the commutator is placed on a Cauchy surface at $t=t_0$, leading to $\langle[\sigma_n(t_0,x),\tilde{\sigma}_n(t_0,x')]\rangle=0$. One could evaluate the commutator of twist operators for other states by using similar method.


\subsection{Commutators of twist operator and its first-order derivative}
In this section, we will first study the commutators of the twist operator and its first-order temporal and spatial derivatives for three examples, for which the expressions of two-point correlators are known. Then, we will use the operator product expansion (OPE) of the twist operator to further discuss the commutators for general states.

\subsubsection{Vacuum state}

Let's first consider the vacuum state as an example. By differentiating equation (\ref{Vacuum state_0}) with respect to time, we obtain the following:
\begin{align}\label{Vacuum_0}
\langle \dot{\sigma}_n(t,x)\tilde{\sigma}_n(t',x')\rangle=\langle \sigma_n(t,x)\tilde{\sigma}_n(t',x')\rangle[\frac{-2h_n}{u-u'-i\epsilon}+\frac{-2\bar h_n}{v-v'-i \epsilon}].
\end{align}
 Eq. (\ref{Vacuum_0}) can be written as
\begin{align}\label{Commutator_Vacuum_01}
 \frac{\langle\dot{\sigma}_n(t,x)\tilde{\sigma}_n(t',x')\rangle}{\langle \sigma_n(t,x)\tilde{\sigma}_n(t',x')\rangle}=-2h_n[i\pi\delta(u-u')+P(\frac{1}{u-u'})+i\pi\delta(v-v')+P(\frac{1}{v-v'})],
\end{align}
by using the Sokhotski-Plemelj theorem 
\begin{align}
\frac{1}{x\mp i\epsilon}=\pm i\pi\delta(x)+P(\frac{1}{x}),
\end{align}
where P denotes the Cauchy principal value.
To calculate the commutator, we need $\langle\tilde{\sigma}_n(t',x')\dot{\sigma}_n(t,x)\rangle$, which is given by $\langle\tilde{\sigma}_n(t',x')\dot{\sigma}_n(t,x)\rangle=\langle\dot{\sigma}_n(t,x)\tilde{\sigma}_n(t',x')\rangle^*$. In Section.\ref{com_twist} we derive that $\langle[\sigma_n(t,x),\tilde{\sigma}_n(t',x')]\rangle=0$, thus $\langle\sigma_n(t,x)\tilde{\sigma}_n(t',x')\rangle^*=\langle\sigma_n(t,x)\tilde{\sigma}_n(t',x')\rangle$. Finally, we have
\begin{align}\label{Commutator_Vacuum_02}
\frac{\langle\tilde{\sigma}_n(t',x')\dot{\sigma}_n(t,x)\rangle}{\langle \sigma_n(t,x)\tilde{\sigma}_n(t',x')\rangle}&=-2h_n[-i\pi\delta(u-u')+P(\frac{1}{u-u'})-i\pi\delta(v-v')+P(\frac{1}{v-v'})].
\end{align}
Combining (\ref{Commutator_Vacuum_01}) and (\ref{Commutator_Vacuum_02}), the commutator is given by
\begin{align}\label{Commutator_Vacuum}
\frac{\langle[\dot{\sigma}_n(t,x),\tilde{\sigma}_n(t',x')]\rangle}{\langle \sigma_n(t,x)\tilde{\sigma}_n(t',x')\rangle}=-4h_n[i\pi\delta(u-u')+i\pi\delta(v-v')].
\end{align}
At the Cauchy surface $t=t_0$, we have
\begin{align}\label{vacuum_com_first}
\frac{\langle[\dot{\sigma}_n(t_0,x),\tilde{\sigma}_n(t_0,x')]\rangle}{\langle \sigma_n(t_0,x)\tilde{\sigma}_n(t_0,x')\rangle}&=-8h_n\pi i \delta(x-x').
\end{align}
For $n\sim 1$, we have
\begin{align}
\frac{\langle[\dot{\sigma}_n(t_0,x),\tilde{\sigma}_n(t_0,x')]\rangle}{\langle \sigma_n(t_0,x)\tilde{\sigma}_n(t_0,x')\rangle}&=\frac{2c}{3}(1-n)\pi i \delta(x-x')+O((n-1)^2),
\end{align}
which is exactly what we expected (\ref{commutator_0}).

On the other hand, we can also examine the commutator that involves the derivative of the twist operator with respect to the spatial coordinate  $x$. The result is
\begin{align}
\frac{\langle[\p_x\sigma_n(t,x),\tilde{\sigma}_n(t',x')]\rangle}{\langle \sigma_n(t,x)\tilde{\sigma}_n(t',x')\rangle}=-4h_n[-i\pi\delta(u-u')+i\pi\delta(v-v')].
\end{align}
Thus, at a certain Cauchy surface $t=t_0$, we have
\begin{align}
\frac{\langle[\p_x\sigma_n(t_0,x),\tilde{\sigma}_n(t_0,x')]\rangle}{\langle \sigma_n(t_0,x)\tilde{\sigma}_n(t_0,x')\rangle}=-4h_n\pi i[-\delta(x'-x)+\delta(x-x')]=0.
\end{align}
This implies that taking the derivative with respect to the spatial coordinate $x$ does not introduce a new imaginary part.

\subsubsection{Thermal state}
Considering the thermal state with temperature $\frac{1}{\beta}$, the two-point correlator of twist operators is 
\begin{align}\label{Thermal state_0}
\langle \sigma_n(t,x)\tilde{\sigma}_n(t',x')\rangle_{\beta}=[\frac{\beta}{\pi}\sinh{(\frac{-\pi(u-u'-i\epsilon)}{\beta}})]^{-2h_n}[\frac{\beta}{\pi}\sinh{(\frac{\pi(v-v'-i\epsilon)}{\beta}})]^{-2\bar h_n}.
\end{align}
Taking the derivative with respect to time, we get
\begin{align}\label{Thermal state_1}
\frac{\langle [\dot{\sigma}_n(t,x),\tilde{\sigma}_n(t',x')]\rangle_{\beta}}{\langle \sigma_n(t,x)\tilde{\sigma}_n(t',x')\rangle_{\beta}}=\frac{-2h_n\cosh{(\frac{\pi(u-u'-i\epsilon)}{\beta}})}{\frac{\beta}{\pi}\sinh{(\frac{\pi(u-u'-i\epsilon)}{\beta}})}+\frac{-2\bar h_n\cosh{(\frac{\pi(v-v'-i\epsilon)}{\beta}})}{\frac{\beta}{\pi}\sinh{(\frac{\pi(v-v'-i\epsilon)}{\beta}})}
\end{align}
The commutator in the thermal state is given by 
\begin{align}
\frac{\langle [\dot{\sigma}_n(t,x),\tilde{\sigma}_n(t',x')]\rangle_{\beta}}{\langle \sigma_n(t,x)\tilde{\sigma}_n(t',x')\rangle_{\beta}}=-4h_n[i\pi\delta(u-u')+i\pi\delta(v-v')].
\end{align}
One could refer to Appendix.(\ref{Details_Thermal}) for the calculation details. 

Then, at Cauchy surface $t=t_0$, the result is
\begin{align}
\frac{\langle [\dot{\sigma}_n(t_0,x),\tilde{\sigma}_n(t_0,x')]\rangle_{\beta}}{\langle \sigma_n(t_0,x)\tilde{\sigma}_n(t_0,x')\rangle_{\beta}}&=-8h_n\pi i\delta(x-x').
\end{align}
When $n\sim1$, we have
\begin{align}
\frac{\langle [\dot{\sigma}_n(t_0,x),\tilde{\sigma}_n(t_0,x')]\rangle_{\beta}}{\langle \sigma_n(t_0,x)\tilde{\sigma}_n(t_0,x')\rangle_{\beta}}&=\frac{2c}{3}(1-n)\pi i\delta(x-x')+O((n-1)^2).
\end{align}
The result is also consistent with (\ref{commutator_0}).

While the commutator involving of the derivative of twist operator with respect to  the spatial coordinate $x$ is 
\begin{align}
\frac{\langle [\p_x{\sigma}_n(t_0,x),\tilde{\sigma}_n(t_0,x')]\rangle_{\beta}}{\langle \sigma_n(t_0,x)\tilde{\sigma}_n(t_0,x')\rangle_{\beta}}&=0.
\end{align}
Once again, the derivation with respect to the spatial coordinate x does not introduce any new imaginary parts.

\subsubsection{States dual to pure AdS$_3$}
Let us further consider the state with holographic dual to the metric satisfying the vacuum Einstein equation in AdS$_3$. By a Conformal mapping $\omega=f(u)$, $\bar\omega=g(v)$, the geometry can be mapped to the Poincare patch. By using Ryu-Takayanagi formula, the holographic EE is given by
\begin{align}
S(t,x;t',x')=\frac{c}{6}\log[-\frac{f(u-i\epsilon)-f(u')}{\sqrt{f'(u-i\epsilon)f'(u')}}][\frac{g(v-i\epsilon)-g(v')}{\sqrt{g'(v-i\epsilon)g'(v')}}].
\end{align}
By definition, we have
\begin{align}
S(t,x;t',x')&=\lim_{n\to 1}\frac{1}{1-n}\log{\langle \Psi|\sigma_n(t,x)\tilde{\sigma}_n(t',x')|\Psi\rangle}\nn\\
\p_t S(t,x;t',x')&=\lim_{n\to 1}\frac{1}{1-n}\frac{\langle \Psi|\dot{\sigma}_n(t,x)\tilde{\sigma}_n(t',x')|\Psi\rangle}{\langle \Psi|\sigma_n(t,x)\tilde{\sigma}_n(t',x')|\Psi\rangle}\nn\\
\end{align}

Taking the derivative with respect to time, we get
\begin{align}
&\lim_{n\to 1}\frac{1}{1-n}\frac{\langle \Psi|\dot{\sigma}_n(t,x)\tilde{\sigma}_n(t',x')|\Psi\rangle}{\langle \Psi|\sigma_n(t,x)\tilde{\sigma}_n(t',x')|\Psi\rangle}\nn\\
&=\frac{c}{6}[\frac{f'(u)}{f(u-i\epsilon)-f(u')}-\frac{f''(u)}{2f'(u)}]+\frac{c}{6}[\frac{g'(v)}{g(v-i\epsilon)-g(v')}-\frac{g''(v)}{2g'(v)}].
\end{align}
Consider the term involving $u$ in the commutator. For $u\neq u'$ we have
\begin{align}
&\frac{c}{6}[\frac{f'(u)}{f(u-i\epsilon)-f(u')}-\frac{f'(u)}{f(u+i\epsilon)-f(u')}]\nonumber \\
&=\frac{f'(u)}{f(u)-f(u')}-\frac{f'(u)}{f(u)-f(u')}=0
\end{align}
If $u\to u'$, the result is divergent. More precisely, we have
\begin{align}
&\frac{c}{6}[\frac{f'(u)}{f(u-i\epsilon)-f(u')}-\frac{f'(u)}{f(u+i\epsilon)-f(u')}]\nn\\
=&\frac{c}{6}[\frac{f'(u')}{f(u')+f'(u')(u-i\epsilon-u')-f(u')}-\frac{f'(u')}{f(u')+f'(u')(u+i\epsilon-u')-f(u')}]\nn\\
=&\frac{c}{6}[\frac{1}{u-u'-i\epsilon}-\frac{1}{u-u'+i\epsilon}]\nn\\
=&\frac{c}{3}\pi i\delta(u-u')
\end{align}
At the Cauchy surface $t = t_0$, the result is
\begin{align}
\lim_{n\to 1}\frac{1}{1-n}\frac{\langle \Psi|[\dot{\sigma}_n(t_0,x),\tilde{\sigma}_n(t_0,x')]|\Psi\rangle}{\langle \Psi|\sigma_n(t_0,x)\tilde{\sigma}_n(t_0,x')|\Psi\rangle}&=\frac{2c}{3}\pi i\delta(x-x').
\end{align}
This result is consistent with our previous expectation (\ref{commutator_0}).

Similarly, we can verify that $\frac{\langle \Psi|[\p_x{\sigma}_n(t_0,x),\tilde{\sigma}_n(t_0,x')]|\Psi\rangle}{\langle \Psi|\sigma_n(t_0,x)\tilde{\sigma}_n(t_0,x')|\Psi\rangle}=0$ also holds true in this context.

\section{Results for general states}\label{section_General}

In the previous sections, we use some examples to show the commutators of twist operators and its temporal and spatial derivative. In this section we would like to discuss the expectation value of the commutators for general states. Some universal results are obtained.

Generally, once knowing operator product expansion (OPE) of two operators, it is possible to evaluate the commutators of the two operators, see, e.g., \cite{Besken:2020snx}. For the twist operators $\sigma_n$ and $\tilde{\sigma}_n$ the OPE\cite{Headrick:2010zt,Calabrese:2010he,Rajabpour:2011pt,Chen:2013kpa} can be expressed as
\bea
\sigma_n(w_1,\bar w_1) \tilde\sigma_n(w_2,\bar w_2)=\sum_{k} C_k (w_1-w_2)^{-2h_{n}+h_k}(\bar w_1-\bar w_2)^{-2\bar h_{n}+\bar h_{ k}} \mathcal{X}_{k},
\eea
where $C_k$ is the OPE coefficent, $\mathcal{X}_k$ are the operators with conformal dimension $h_k\geq0$ and $\bar h_k\geq0$.  With Wick rotation, we have 
\begin{align}\label{General states_0}
\langle \Psi|\sigma_n(t,x)\tilde{\sigma}_n(t',x')|\Psi\rangle=\langle \sigma_n(t,x)\tilde{\sigma}_n(t',x')\rangle\sum_{k} a_k [-(u-u'-i\epsilon)]^{h_k}[v-v'-i\epsilon]^{\bar h_{ k}} ,
\end{align}
where $a_k$ depends on OPE coefficients and the expectation values of the operators in the state $|\psi\rangle$ and $\langle \sigma_n(t,x)\tilde{\sigma}_n(t',x')\rangle$ denotes the contribution for the vacuum state. When
 the length of the interval is short enough, we can consider that $a_k$ are constants which are independent of the coordinates. Throughout the remainder of this section, we adopt the short interval assumption. 

\subsection{Universal result for commutator involving first-order temporal derivative}

In this section we would like to calculate $\frac{\langle \Psi|[\dot{\sigma}_n(t_0,x),\tilde{\sigma}_n(t_0,x')]|\Psi\rangle}{\langle \Psi|\sigma_n(t_0,x)\tilde{\sigma}_n(t_0,x')|\Psi\rangle}$ for most general states. Taking the logarithm of Eq.(\ref{General states_0}), we have
\begin{align}\label{G_log}
&\log\langle \Psi|\sigma_n(t,x)\tilde{\sigma}_n(t',x')|\Psi\rangle\nn\\
=&\log\langle \sigma_n(t,x)\tilde{\sigma}_n(t',x')\rangle+\log\sum_{k} a_k [-(u-u'-i\epsilon)]^{h_k}[v-v'-i\epsilon]^{\bar h_{ k}} \nn\\
=&\log\langle \sigma_n(t,x)\tilde{\sigma}_n(t',x')\rangle+\log a_0+\sum_{k\neq0} \frac{a_k}{a_0} [-(u-u'-i\epsilon)]^{h_k}[v-v'-i\epsilon]^{\bar h_{ k}},
\end{align}
where $h_{0}=0$ accounts for the contribution of the identity operator, and in the following we set $a_0=1$. In the second step, we expand the logarithm under the assumption that the series is convergent. 

Taking the derivative of the above result with respect to $t$ gives
\begin{align}
&\frac{\langle \Psi|\dot{\sigma}_n(t,x)\tilde{\sigma}_n(t,x')|\Psi\rangle}{\langle \Psi|\sigma_n(t,x)\tilde{\sigma}_n(t,x')|\Psi\rangle}=\p_t\log\langle \Psi|\sigma_n(t,x)\tilde{\sigma}_n(t',x')|\Psi\rangle\nn\\
=&\frac{\langle \dot{\sigma}_n(t,x)\tilde{\sigma}_n(t,x')\rangle}{\langle \sigma_n(t,x)\tilde{\sigma}_n(t,x')\rangle}+\sum_{k\neq0} a_k [-(u-u'-i\epsilon)]^{h_k}[v-v'-i\epsilon]^{\bar h_{ k}}[\frac{h_k}{u-u'-i\epsilon}+\frac{\bar h_k}{v-v'-i \epsilon}],
\end{align}
Taking the result at a certain Cauchy surface $t=t_0$ and taking the commutator, we get
\begin{align}\label{G_1}
&\frac{\langle \Psi|[\dot{\sigma}_n(t_0,x),\tilde{\sigma}_n(t_0,x')]|\Psi\rangle}{\langle \Psi| \sigma_n(t_0,x)\tilde{\sigma}_n(t_0,x')|\Psi\rangle}\nn\\
=&\frac{\langle [\dot{\sigma}_n(t_0,x),\tilde{\sigma}_n(t_0,x')]\rangle}{\langle  \sigma_n(t_0,x)\tilde{\sigma}_n(t_0,x')\rangle}+\sum_{k\neq0} a_k h_k (x-x')^{2h_k}4\pi i\delta(x-x')\nn\\
=&-8h_n\pi i \delta(x-x')
\end{align}
Using the properties of the Dirac delta function  $f(x)\delta(x) = f(0)\delta(x)$, we can derive in the second step that for $h_k > 0$, the terms $(x-x')^{2h_k}\delta(x-x')$ are vanishing. The above result is also consisent with Eq.(\ref{commutator_0}).

Eq.(\ref{G_1}) demonstrates that the expectation value of the commutator involving the twist operator and its first-order temporal derivative is universal, which depends only on the central charge $c$ and is independent of the states. This result is consistent with the examples discussed in the previous section. By using Eq.(\ref{Im_0}), we can observe that the imaginary components of the timelike EE for the vacuum state, thermal state, and states with holographic duals are directly linked to these commutators. Our findings explain why the imaginary term is the universal value $\frac{i \pi c}{6}$. In fact, we can anticipate that for any state, the imaginary term of the timelike EE will include a contribution of $\frac{i \pi c}{6}$.

\subsection{The derivative of the commutator with respect to the spatial coordinate $x$}
On the other hand, we can consider the commutator with respect to the derivative of the spatial coordinate $x$
\begin{align}
&\frac{\langle \Psi|\p_x\sigma_n(t,x)\tilde{\sigma}_n(t,x')|\Psi\rangle}{\langle \Psi|\sigma_n(t,x)\tilde{\sigma}_n(t,x')|\Psi\rangle}=\p_x\log\langle \Psi|\sigma_n(t,x)\tilde{\sigma}_n(t',x')|\Psi\rangle\nn\\
=&\frac{\langle \p_x\sigma_n(t,x)\tilde{\sigma}_n(t,x')\rangle}{\langle \sigma_n(t,x)\tilde{\sigma}_n(t,x')\rangle}+\sum_{k\neq0} a_k [-(u-u'-i\epsilon)]^{h_k}[v-v'-i\epsilon]^{\bar h_{ k}}[\frac{-h_k}{u-u'-i\epsilon}+\frac{\bar h_k}{v-v'-i \epsilon}],
\end{align}

At a certain Cauchy surface $t=t_0$, we have
\begin{align}
&\frac{\langle \Psi|[\p_x\sigma_n(t_0,x),\tilde{\sigma}_n(t_0,x')]|\Psi\rangle}{\langle\Psi| \sigma_n(t_0,x)\tilde{\sigma}_n(t_0,x')|\Psi\rangle}\nn\\
=&0+\sum_{k\neq0} a_k [x-x'+i\epsilon]^{h_k}[x-x'-i\epsilon]^{h_{ k}}[\frac{h_k}{x-x'+i\epsilon}+\frac{ h_k}{x-x'-i\epsilon}]\nn\\
-&0-\sum_{k\neq0} a_k [x-x'-i\epsilon]^{h_k}[x-x'+i\epsilon]^{h_{ k}}[\frac{h_k}{x-x'-i\epsilon}+\frac{ h_k}{x-x'+i\epsilon}]\nn\\
=&0
\end{align}
Thus for general states in 2D CFT, the first derivative of its commutator with respect to the spatial coordinate $x$ is 0. This implies that the derivative of the EE with respect to spatial coordinate $\p_x S(t_0,x;t_0,x')$ is a real number, and it does not contribute to a new imaginary part.

\subsection{Results for commutators involving higher order derivative}\label{section_general_results}

We have derived a universal result regarding the first-order derivative of the commutator for general states. However, this indicates that the first-order derivative can only yield information analogous to that of the vacuum state. For more general states, the imaginary parts of the timelike EE should also depend on the specific details of those states, as we will demonstrate below. Additionally, we will show that it is essential to consider commutators that involve higher-order temporal derivatives of the twist operators. Furthermore, we will demonstrate that the imaginary part of the timelike EE can be related to the $m$-th derivatives of spacelike EE or the commutators of twist operators. Specifically, we will construct a formula analogous to equation (\ref{Im_0}).


Based on the results from the previous section, we have
\begin{align}
\p_t S(t_0,x;t_0,x')=\lim_{n\to 1}\frac{-4h_n\pi i \delta(x-x')}{1-n}=\frac{c}{3}\pi i\delta(x-x')
\end{align}
For higher-order derivatives, we note that the even-order derivatives are always real. For instance, considering the second derivative, we have
\begin{align}\label{S_G_D2}
&\p^2_t S(t_0,x;t_0,x')\nn\\
=&\lim_{n\to 1}\frac{1}{1-n}\{ -2h_n[\frac{-1}{(x-x'-i\epsilon)^2}+\frac{-1}{(x-x'+i\epsilon)^2}]\nn\\
+&\sum_{k\neq0} a_k [C^0_2 h_k(h_k-1)+C^1_2 h_k^2+C^2_2 h_k(h_k-1)] (x-x')^{2h_k-2}\}\nn\\
=&-\frac{c}{3}(x-x')^{-2}+\sum_{k\neq0} \tilde{a}_k 2h_k(2h_k-1) (x-x')^{2h_k-2},
\end{align}
where $\tilde a_k:=\lim_{n\to 1}\frac{1}{1-n} a_k$, $C^k_m:=\frac{m!}{k!(m-k)!}$.
This term does not provide any additional information regarding the new imaginary part of the timelike separation. More generally, for higher-order even derivatives, all yield real numbers, as indicated by the aforementioned Eq.(\ref{S_G_D2}). Therefore, we are particularly interested in the odd-order derivatives.

For the third-order derivative, we have
\begin{align}\label{third_derivative}
&\p^3_t S(t_0,x;t_0,x')\nn\\
=&\lim_{n\to 1}\frac{1}{1-n}\{ -4h_n[\frac{1}{(x-x'-i\epsilon)^3}-\frac{1}{(x-x'+i\epsilon)^3}]\nn\\
+&\sum_{k\neq0} a_k [C^0_3 h_k(h_k-1)(h_k-2)+C^1_3 h_k^2(h_k-1)] (x-x')^{2h_k-2}[\frac{1}{x-x'-i\epsilon}-\frac{1}{x-x'+i\epsilon}]\}\nn\\
=&\frac{c}{3}\pi i \p^2_x\delta(x-x')\nn \\
+&\sum_{0<h_k<1} \tilde a_k [C^0_3 h_k(h_k-1)(h_k-2)+C^1_3 h_k^2(h_k-1)] (x-x')^{2(h_k-1)}2\pi i \delta(x-x').
\end{align}
 As we expected, the third derivative of the EE with respect to time $t$ at the Cauchy surface $t = t_0$ will be non-vanishing with $0<h_k<1$. In fact, as we can see from Eq.(\ref{third_derivative}), this term is divergent. We will demonstrate below that this term is crucial for producing the correct imaginary part of the timelike EE. Notably, since each term in the coefficient contains a factor of $h_k-1$, this term is vanishing when $h_k=1$.

In general, we have
\begin{align}\label{G_high}
&\p^{(2m+1)}_t S(t_0,x;t_0,x')=\frac{c}{3}\pi i \p^{2m+1}_x\delta(x-x')\nn \\
+&\sum_{0<h_k\leq m} \tilde a_k [C^0_{2m+1} h_k(h_k-1)...(h_k-2m)+C^1_{2m+1} h_k^2(h_k-1)...(h_k-2m+1)\nn\\
+&C^2_{2m+1} h_k^2(h_k-1)^2(h_k-2)...(h_k-2m+2)+...\nn\\
+&C^m_{2m+1} h_k^2(h_k-1)^2...(h_k-m+1)^2(h_k-m)] (x-x')^{2(h_k-m)}2\pi i \delta(x-x')\nn\\
=&\frac{c}{3}\pi i \p^{2m+1}_x\delta(x-x')+\sum_{0<h_k\leq m} \tilde a_k G^{(2m+1)}_k (x-x')^{2(h_k-m)}2\pi i \delta(x-x')\nn\\
=&\frac{c}{3}\pi i \p^{2m+1}_x\delta(x-x')+\sum_{\tilde{k}} \tilde a_{\tilde{k}} G^{(2m+1)}_{\tilde{k}} (x-x')^{2(h_{\tilde{k}}-m)}2\pi i \delta(x-x'),
\end{align}
where we denote the term with fractional conformal dimension $h_k$ within $\{h_k\}$ as $\{ h_{\tilde k}\}$, and we define
\begin{align}
G^{(2m+1)}_k:=C^0_{2m+1} h_k(h_k-1)...(h_k-2m)+...+C^m_{2m+1} h_k^2(h_k-1)^2...(h_k-m+1)^2(h_k-m).
\end{align}
Let us take $m=[h_k]$, where $[x]$ denotes the ceiling function of $x$. For integer values of $h_k$, since each term in $G^{(2m+1)}_k$ contains a factor of $(h_k-1)...(h_k-m)$, it follows that $G^{(2m+1)}_k=0$. Therefore, in the final step of the above equation, the terms with integers conformal dimension will be vanishing, leaving us with only the fractional terms $\{ h_{\tilde k}\}$. These terms are imaginary, which will be important to construct the formula of imaginary part of timelike EE.

Therefore, we know that the $(2m+1)$th derivative of the EE with respect to time $t$ will include the information of the operators with conformal dimension $h_k$, where $h_k$ is a fraction with $m=[h_k]$. Provided that we choose $m$ to be sufficiently large, it will encompass all the contributions of the operators with conformal dimension $h_k$.

In fact, operators with integer values of $h_k$ will not provide any new contributions to the imaginary part of timelike EE. In contrast, operators with fractional $h_k$ will contribute significantly. In the next section, we will analyze how to modified the relation between timelike and spacelike EE by including all the possible contributions from the operators with conformal dimension $h_k$.

\subsection{General formula for the imaginary part of timelike EE}
We know that, for general states, by using (\ref{G_log}) the EE is
\begin{align}\label{S_G_0}
S(t,x;t',x')=S_0(t,x;t',x')+\sum_{k\neq0} \tilde a_k [-(u-u'-i\epsilon)]^{h_k}[v-v'-i\epsilon]^{\bar h_{ k}}.
\end{align}
For spacelike separation $(t,x;t',x')$, we have
\begin{align}
S(t,x;t',x')=S_0(t,x;t',x')+\sum_{k\neq0} \tilde a_k \Delta s^{2h_k},
\end{align}
where we assume $h_k=\bar h_k$.
And for timelike separation $(t,x;t',x')$, we have
\begin{align}
S(t,x;t',x')=S_0(t,x;t',x')+\sum_{k\neq0} \tilde a_k  (-1)^{h_k}T_0^{2h_k},
\end{align}
where we define $T_0^{2}=-\Delta s^{2}=(t-t')^2-(x-x')^2$.
Regarding the timelike separation, we have observed that terms with integer values of $h_k$ do not contribute additional imaginary parts.  it is only the terms with fractional values of $h_k$ that yield new imaginary components. Let us denote all fractional terms within $\{h_k\}$ as $\{ h_{\tilde k}\}$. Subsequently, the imaginary part of timelike EE is expressed as
\begin{align}\label{S_G_I}
\text{Im} S(t,x;t',x')=\frac{i\pi c}{6}+\sum_{\tilde k} \text{Im}[(-1)^{h_{\tilde k}}] \tilde a_{\tilde k} T_0^{2h_{\tilde k}},
\end{align}

By employing the results outlined in Section \ref{section_general_results}, we aim to establish a relationship analogous to Eq.(\ref{Im_0}) for general states. To accomplish this, we begin with Eq.(\ref{G_high}) and define  the function $S_{\tilde{k}}(t_0,x;t_0,x')$ satisfying
\begin{align}\label{function_s_tilde}
&\p^{(2m+1)}_t S_{\tilde{k}}(t_0,x;t_0,x'):=\tilde a_{\tilde{k}} G^{(2m+1)}_{\tilde{k}} (x-x')^{2(h_{\tilde{k}}-m)}2\pi i \delta(x-x').
\end{align}
It is obvious that by definition we have
\bea
\sum_{\tilde{k}}\p^{(2m+1)}_t S_{\tilde{k}}(t_0,x;t_0,x')= \p^{(2m+1)}_t S(t_0,x;t_0,x')-\frac{c}{3}\pi i \p^{2m+1}_x\delta(x-x').
\eea
It is noteworthy that, from Eq. (\ref{S_G_I}), the additional imaginary parts are exclusively associated with operators that have fractional conformal dimensions $h_{\tilde{k}}$.  More specifically, these contributions depend on the parameters $\tilde{a}_{\tilde{k}}$ and $h_{\tilde{k}}$. The function $S_{\tilde{k}}(t_0,x;t_0,x')$ defined in (\ref{function_s_tilde}) also incorporates these parameters. Therefore, we expect that the relation in Eq. (\ref{Im_0}) can be generalized to include contributions from $S_{\tilde{k}}(t_0,x;t_0,x')$ to more general states.

The formula Eq.(\ref{Im_0}) should be modified by including more terms associated with $S_{\tilde{k}}(t_0,x;t_0,x')$. Let us denote the corrections by $Im \Delta S(t,x;t',x')$. By considering the corrections from  operators with fractional conformal dimension, we propose a correction to Eq.(\ref{Im_0})
\begin{align}\label{S_G}
&2\text{Im}\Delta S(t,x;t',x')\nn \\
=&\sum_{\tilde{k}}\frac{[\text{Im}e^{i\pi h_{\tilde{k}}}]e^{\frac{-i\pi}{2}}T_0^{2h_{\tilde{k}}}}{4\pi G^{(2m+1)}_{\tilde{k}} }\Big[\int_{-u'}^{v'}\text{d}y (-u-y)^{-2(h_{\tilde{k}}-m)}\p^{(2m+1)}_{t'}S_{\tilde{k}}(t_0,-u;t_0,y)\nn\\
+&\int_{-u'}^{v'}\text{d}y (v-y)^{-2(h_{\tilde{k}}-m)}\p^{(2m+1)}_{t'}S_{\tilde{k}}(t_0,v;t_0,y)\nn \\
+&\int_{-u}^{v}\text{d}y (y+u')^{-2(h_{\tilde{k}}-m)}\p^{(2m+1)}_{t}S_{\tilde{k}}(t_0,y;t_0,-u')\nn \\
+&\int_{-u}^{v}\text{d}y (y-v')^{-2(h_{\tilde{k}}-m)}\p^{(2m+1)}_{t}S_{\tilde{k}}(t_0,y;t_0,v')-\text{h.c.}\Big]
\end{align}
where $m=[h_k]$ is the result of applying the ceiling function to the maximum value in the set $\{h_k\}$, and $h.c.$ denotes the complex conjugate. Thus the full imaginary part of timelike EE is expressed as follows
\bea
\text{Im} S_f(t,x;t',x')=\text{Im} S(t,x;t',x')+\text{Im} \Delta S(t,x;t',x'),
\eea
where $\text{Im} S(t,x;t',x')$ is given by Eq.(\ref{Im_0}) and contributes the universal term $\frac{i\pi  c}{6}$, while $\text{Im} \Delta S(t,x;t',x')$ accounts for the contributions from all operators with fractional conformal dimensions.

As a simple verification of Eq. (\ref{S_G}), one can demonstrate that it becomes zero for the spacelike separation between the points $(t,x)$ and $(t',x')$. As discussed in Section \ref{section_2},  in the scenario depicted in Fig. \ref{separation} (c), the integral results from (\ref{S_G}) yield a value of zero. While, in the case illustrated in Fig. \ref{separation} (d), the contributions cancel each other out. This aligns with our expectation that the imaginary part of the spacelike EE vanishes.

For timelike separation between the points $(t,x)$ and $(t',x')$, according to our formula (\ref{S_G}), one could check that the imaginary part of the EE is $\text{Im} S_f(t,x;t',x')=\frac{i\pi c}{6}+\sum_{\tilde k} \text{Im}[(-1)^{h_{\tilde k}}] \tilde a_{\tilde k} T_0^{2h_{\tilde k}}$, which is consistent with the timelike EE (\ref{S_G_I}). As we have noted that the universal term $\frac{i\pi c}{6}$ come from $\text{Im} S(t,x;t',x')$ (\ref{Im_0}). Specifically, when all elements in the set of $\{h_k\}$ are integers, the imaginary part is $\text{Im} S(t,x;t',x')=\frac{i\pi c}{6}$, meaning that the integer terms do not contribute to a new imaginary part.

Let us briefly summarize the results of this section as follows. Suppose that the EE can be evaluated using the OPE of twist operators. For general states, the imaginary part of the timelike EE is contributed solely by the identity operator and operators with fractional conformal dimensions $\{h_{\tilde k}\}$. The identity operator contributes a universal term of $\frac{i\pi c}{6}$, while the operators with fractional conformal dimension $h_{\tilde k }$ contribute  $\sum_{\tilde k} \text{Im}[e^{i\pi h_{\tilde k}}] \tilde a_{\tilde k} T_0^{2h_{\tilde k}}$. 

In the general case, the imaginary part of the timelike EE can also be expressed in terms of the spacelike EE and its derivatives on a specific Cauchy surface at $t = t_0$, as shown in Eq.(\ref{S_G}). We have observed that terms involving even-order derivatives, such as $\frac{1}{4}\int_{-u}^{v}\text{d}y \int_{-u'}^{v'}\text{d}y'\p_{t}\p_{t'}S(t_0,y;t_0,y')$, do not contribute a new imaginary part but instead provide real part contributions. The novel imaginary part arises solely from first-order derivatives and higher-order odd derivatives. While the even-order derivatives are important for the real part of the timelike EE, both even-order and odd-order derivatives will be essential for extending Eq. (\ref{S_separation}) to more general states. We plan to explore this question in the near future.

\section{Higher dimensional example}\label{section_Higher dimension}

We consider that $\text{AdS}_{d+1}$ possesses the metric
\begin{align}\label{metric}
ds^2=R^2\frac{-dt^2+dz^2+dx^2+\sum_{i=1}^{d-2}dy_i^2}{z^2},
\end{align}
where R is the radius of $\text{AdS}$ space. The subsystem  $A$ is specified by the boosted strip defined by
\begin{align}
&-\frac{\Delta x}{2}<x<\frac{\Delta x}{2},-\frac{\Delta t}{2}<t<\frac{\Delta t}{2},x=\frac{\Delta x}{\Delta t} t;\nn\\
&-\frac{L}{2}<y_i<\frac{L}{2},i=1,2,...,d-2,
\end{align}
where $L$ the IR cut-off on the coordinate $y_i$. 

In fact, we can regard $A$ as a region bounded by two hypersurfaces with constant coordinates $(t,x)$ and $(t', x')$ with $x-x'=\Delta x$ and $t-t'=\Delta t$.
One could extend the $A$ to timelike region. For instance, in the case where $x=x'=0$, it represents a timelike strip on the hypersurface with $x=0$.  According to our previous convention, we still define $u = t - x$, $u' = t' - x'$ and $v = t + x$, $v' = t' + x'$. For region $A$ the holographic EE is given by \cite{Kusuki:2017}\cite{Heller:2024}
\begin{align}\label{S_Higher dimension}
S_A=S(t,x;t',x')&=\frac{L^{d-2}R^{d-1}}{2G_N}[\frac{1}{d-2}(\frac{1}{\delta})^{d-2}-k_d\frac{1}{\Delta s^{d-2}}]\nn\\
&=\frac{L^{d-2}R^{d-1}}{2G_N}[\frac{1}{d-2}(\frac{1}{\delta})^{d-2}-k_d\frac{(-i)^{d-2}}{T_0^{d-2}}],
\end{align}
where we defined
\begin{align}
k_d&=\frac{\pi^{\frac{d-1}{2}2^{d-2}}}{d-2}(\frac{\Gamma(\frac{d}{2(d-1)})}{\Gamma(\frac{1}{2(d-1)})})^{d-1}\nn\\
T_0^2&:=-\Delta s^2=(u-u'-i\epsilon)(v-v'-i\epsilon)=(t-t')^2-(x-x')^2.
\end{align}
There are two distinct scenarios for high-dimensional timelike EE in odd and even dimensions, which are
\begin{align}\label{S_t_hd}
S(t,x;t',x')=\left\{ 
    \begin{array}{lc}
        \frac{L^{d-2}R^{d-1}}{2G_N}[\frac{1}{d-2}(\frac{1}{\delta})^{d-2}-k_d\frac{-i(-1)^\frac{d-3}{2}}{T_0^{d-2}}] & \text{d is odd}, \\
        \frac{L^{d-2}R^{d-1}}{2G_N}[\frac{1}{d-2}(\frac{1}{\delta})^{d-2}-k_d\frac{(-1)^\frac{d-2}{2}}{T_0^{d-2}}] & \text{d is even}.\\
    \end{array}
\right.
\end{align}
This indicates that the high-dimensional timelike EE for the strip has an imaginary part only when $d$ is odd, while it remains purely real in even dimensions.

Based on the discussion in 2-dimensional CFTs, the imaginary part of the timelike EE is related to the commutator of twist operators that involve temporal derivatives. To proceed, let us take the derivative of equation (\ref{S_Higher dimension}) with respect to time $t$, yielding
\begin{align}
&\lim_{n\to 1}\frac{1}{1-n}\frac{\langle \Psi|\dot{\sigma}_n(t,x)\tilde{\sigma}_n(t',x')|\Psi\rangle}{\langle \Psi|\sigma_n(t,x)\tilde{\sigma}_n(t',x')|\Psi\rangle}=\p_{t} S(t,x;t',x')\nn\\
=&\frac{L^{d-2}R^{d-1}}{2G_N}\frac{k_d(d-2)}{2}[-(u-u'-i\epsilon)(v-v'-i\epsilon)]^{-\frac{d-2}{2}}(\frac{1}{u-u'-i\epsilon}+\frac{1}{v-v'-i\epsilon})
\end{align}
On a certain Cauchy surface $t=t_0$, we would obtain the following relation 
\begin{align}
&(x-x')^{(d-2)}\p_{t} S(t_0,x;t_0,x')\nn\\
=&\frac{L^{d-2}R^{d-1}}{2G_N}k_d(d-2)\pi i\delta(x-x'),
\end{align}
where $\p_{t} S(t_0,x;t_0,x')$ should be understood as $\p_{t} S(t,x;t',x')|_{t\to t_0,t'\to t_0}$. The result can be rewritten as the commutators of twist operators, that is
\begin{align}\label{commutator_hd}
&\lim_{n\to 1}\frac{1}{1-n}(x-x')^{(d-2)}\frac{\langle \Psi|[\dot{\sigma}_n(t_0,x),\tilde{\sigma}_n(t_0,x')]|\Psi\rangle}{\langle \Psi|\sigma_n(t_0,x)\tilde{\sigma}_n(t_0,x')|\Psi\rangle}\nn\\
=&\frac{L^{d-2}R^{d-1}}{G_N}k_d(d-2)\pi i\delta(x-x').
\end{align}

Interestingly, the commutators of twist operators for the strip exhibit properties similar to those of two-dimensional CFTs (\ref{vacuum_com_first}). As indicated in (\ref{Im_0}), the imaginary part of the timelike entanglement entropy (EE) is closely related to the commutators of the twist operator and its first temporal derivative. We hypothesize that a similar relationship exists for the strip, analogous to that of two-dimensional CFTs. One can evaluate the imaginary part of the timelike EE for the strip by utilizing (\ref{S_Higher dimension}) and the definition
\bea\label{Imaginary_strip}
2\text{Im} S(t,x;t',x')=S(t,x;t',x')-S(t,x;t',x')^*.
\eea
For any separation $(t,x;t',x')$, we can conclude that there is a relation analogous to (\ref{Im_0}), which is expressed as follows
\begin{align}\label{Im_hd}
&2\text{Im} S(t,x;t',x')\nn\\
&=\frac{[(-1)^{d-2}-1]i^{d-3}}{T_0^{d-2}(d-2)\pi}\frac{1}{4}\int_{-u'}^{+v'}(-u-y)^{(d-2)}\lim_{n\to 1}\frac{1}{1-n}\frac{\langle \Psi|[\dot{\sigma}_n(t_0,-u),\tilde{\sigma}_n(t_0,y)]|\Psi\rangle}{\langle \Psi|\sigma_n(t_0,-u)\tilde{\sigma}_n(t_0,y)|\Psi\rangle} \text{d}y\nn\\
&+\frac{[(-1)^{d-2}-1]i^{d-3}}{T_0^{d-2}(d-2)\pi}\frac{1}{4}\int_{-u'}^{+v'}(v-y)^{(d-2)}\lim_{n\to 1}\frac{1}{1-n}\frac{\langle \Psi|[\dot{\sigma}_n(t_0,v),\tilde{\sigma}_n(t_0,y)]|\Psi\rangle}{\langle \Psi|\sigma_n(t_0,v)\tilde{\sigma}_n(t_0,y)|\Psi\rangle} \text{d}y\nn\\
&+\frac{[(-1)^{d-2}-1]i^{d-3}}{T_0^{d-2}(d-2)\pi}\frac{1}{4}\int_{-u}^{+v}(y-(-u'))^{(d-2)}\lim_{n\to 1}\frac{1}{1-n}\frac{\langle \Psi|[\dot{\sigma}_n(t_0,y),\tilde{\sigma}_n(t_0,-u')]|\Psi\rangle}{\langle \Psi|\sigma_n(t_0,y)\tilde{\sigma}_n(t_0,-u')|\Psi\rangle} \text{d}y\nn\\
&+\frac{[(-1)^{d-2}-1]i^{d-3}}{T_0^{d-2}(d-2)\pi}\frac{1}{4}\int_{-u}^{+v}(y-v')^{(d-2)}\lim_{n\to 1}\frac{1}{1-n}\frac{\langle \Psi|[\dot{\sigma}_n(t_0,y),\tilde{\sigma}_n(t_0,v')]|\Psi\rangle}{\langle \Psi|\sigma_n(t_0,y)\tilde{\sigma}_n(t_0,v')|\Psi\rangle} \text{d}y.
\end{align}
One can directly verify the above formula using (\ref{Imaginary_strip}) and (\ref{commutator_hd}). As we mentioned, the imaginary part of the timelike EE for the strip vanishes in even dimensions, and the above formula automatically yields a zero result for even dimensions. In contrast, for odd dimensions, it produces the correct imaginary part as shown in (\ref{S_t_hd}). This can be seen as a non-trivial check of the above formula. It is also worth noting that the expression (\ref{Im_hd}) will vanish for spacelike separations in both odd and even dimensions. This observation is consistent with the fact that the EE in spacelike separations does not contain an imaginary part.

Remarkably, the formula (\ref{Im_hd}) reduces to the two-dimensional case (\ref{Im_0}) in the limit $d\to 2$. This can be verified by observing that
\begin{align}
\lim_{d\to 2}\frac{2i[(-i)^{d-2}-1]}{T_0^{d-2}(d-2)\pi}
=1.
\end{align}
This once again supports the validity of the formula (\ref{Im_hd}). 

On the other hand, we can also express Eq.(\ref{Im_hd}) in terms of time partial derivatives of the spacelike EE, analogous to Eq.(\ref{S_G}). The result is
\begin{align}\label{Im_hd_S}
&2\text{Im} S(t,x;t',x')\nn\\
&=\frac{[(-1)^{d-2}-1]i^{d-3}}{T_0^{d-2}(d-2)\pi}\frac{1}{4}\Big[\int_{-u'}^{+v'}(-u-y)^{(d-2)}\p_{t'}S(t_0,-u;t_0,y) \text{d}y\nn\\
&\phantom{=\frac{[(-1)^{d-2}-1]i^{d-3}}{T_0^{d-2}(d-2)\pi}\frac{1}{4}\Big[}+\int_{-u'}^{+v'}(v-y)^{(d-2)}\p_{t'}S(t_0,v;t_0,y)\text{d}y\nn \\
&\phantom{=\frac{[(-1)^{d-2}-1]i^{d-3}}{T_0^{d-2}(d-2)\pi}\frac{1}{4}\Big[}+\int_{-u}^{+v}(y+u')^{(d-2)} \p_{t}S(t_0,y;t_0,-u')\text{d}y\nn \\
&\phantom{=\frac{[(-1)^{d-2}-1]i^{d-3}}{T_0^{d-2}(d-2)\pi}\frac{1}{4}\Big[}+\int_{-u}^{+v}(y-v')^{(d-2)}\p_{t}S(t_0,y;t_0,v')\text{d}y-h.c.\Big],
\end{align}
where $h.c.$ denotes the complex conjugate.
We will discuss the possible extensions and applications of the formula (\ref{Im_hd}) in the discussion section.


\section{Commutators and spectra of reduced density matrix}\label{section_spectra}
We consider the R\'enyi entropy for one interval between $(t,x)$ and $(t',x')$. The two points can be spacelike or timelike. 
For  vacuum state of 2-dimensional CFTs the R\'enyi entropy is given by
\bea\label{Renyivacuum}
S_n(t,x;t',x')=\frac{c}{12}(1+\frac{1}{n})\log \frac{-(u-u'-i\epsilon)(v-v'-i\epsilon)}{\delta^2},
\eea
where $u:=t-x$,$u':=t'-x'$ and $v:=t+x$, $v':=t'-x'$, $\epsilon$ is a positive constant. One can obtain the above result by analytical continuation of the correlator of twist operators. An obvious property is 
\bea
S_n(t,x;t',x')=S_n(t',x';t,x)^*.
\eea
One could reconstruct the spetra of $\rho_A$ once knowing R\'enyi entropy for all the $n$.
Let us consider the separation of two points are spacelike. The  density matrix $\rho_A$ is still Hermitian, thus the eigenvalues $\lambda_i$ are still positive numbers $\lambda_i \in (0,1)$. Denote the maximal eigenvalue as $\lambda_m$, which is given by $b:=-\log \lambda_m=\lim_{n\to \infty}S_n$. Thus the R\'enyi entropy (\ref{Renyivacuum}) can be written as $S_n(t,x;t',x')=(1+\frac{1}{n})b$. 
 From the form of R\'enyi entropy we expect the eigenvalues of $\rho_A$ should be $\lambda_i=\lambda_i(t,x;t',x')$ and satisfy the similar property $\lambda_i(t,x;t',x')=\lambda_i(t',x';t,x)^*$. A special case is the maximal eigenvalue $\lambda_m(t,x;t',x'):=e^{-b(t,x;t',x')}$. One could check that $b$ as well as $\lambda_m$ are real for spacelike separation.

 For the spacelike separation the eigenvalues should be real that is $\lambda_i(t,x;t',x')=\lambda_i(t',x';t,x)$
Using R\'enyi entropy one could evaluate the following functions,
\bea
&&\mP(\lambda):= \sum_i \delta(\lambda-\lambda_i),\nn\\
&&\mP_{\alpha_J}(\lambda):=\sum_i \frac{\p \lambda_i}{\p \alpha_J}\delta(\lambda-\lambda_i),
\eea
where $\alpha_J$ is arbitrary parameter.
$\mP$ is the density of eigenvalue, $\mP_{\alpha_J}/\mP$ can be seen as the average value of the derivative of eigenvalue at $\lambda$ with respect to $\alpha_J$. By definition we would have the following relation
\bea
\frac{\p \mP}{\p \alpha_J}=-\frac{\p \mP_{\alpha_J}}{\p \lambda}.
\eea

In this paper we are interested in the case $\alpha_J=t$. The two functions $\mP$ and $\mP_t$ can be computed as follows,
\bea
&&\mP(\lambda)=\lambda^{-1}\mathcal{L}^{-1}\left[e^{nb+(1-n)S_n}\right],\nn \\
&&\mP_t(\lambda)=\mathcal{L}^{-1}\left[ \frac{1-n}{n}e^{nb+(1-n)S_n }\frac{\p S_n}{\p t}\right].
\eea
For the vacuum state the result is
\bea
&&\mP(\lambda)=\frac{1}{\lambda}\left[\frac{\sqrt{b} I_1\left(2 \sqrt{b s}\right)}{\sqrt{s}}+\delta (s)\right],\nn \\
&&\mP_t(\lambda)=-\frac{\p b}{\p t}\left[\frac{(b-s) I_1\left(2 \sqrt{bs}\right)}{\sqrt{b s}}+\delta (s)\right],
\eea
where 
\bea
\frac{\p b}{\p t}=\frac{c}{12}\left( \frac{1}{v-v'-i\epsilon}+\frac{1}{u-u'-i\epsilon}\right).
\eea

We should understand $b$ as $b(0,x;0,x')$, but understand $\frac{\p b}{\p t}$ as $\frac{\p b(t,x;t',x')}{\p t}|_{t=0,t'=0}$. This is why $b$ is real and $\frac{\p b}{\p t}$ is complex.

For the spacelike separation one could find $\mP=\mP^*$. However, for the function $\mP_t$ is complex, the imaginary part is
\bea\label{imaginarypart_vacuum}
\text{Im} \mP_t=\frac{1}{2}\left(\mP_t-\mP_t^*\right)=-\frac{1}{2}\left(\frac{\p b}{\p t}-\frac{\p b^*}{\p t} \right)\left[\frac{(b-s) I_1\left(2 \sqrt{bs}\right)}{\sqrt{b s}}+\delta (s)\right].
\eea

By definition the time derivative of $tr\rho_A^n$ is
\bea
\frac{\p }{\p t} tr \rho_A^n=
n\sum_i\frac{\p \lambda_i}{\p t}\lambda_i^{n-1}=n\int_{0}^{\lambda_m} d\lambda \lambda^{n-1}\mP_t(\lambda)
\eea
In QFTs one could evaluate $tr\rho_A^n$ by twist operator formalism, that is 
\bea
tr\rho_A^n=\langle \sigma_n(t,x)\tilde{\sigma}_n(t',x')\rangle,
\eea
where we taking the analytical continuation to Lorentzian coordinate. Thus we find
\bea
\langle \dot{\sigma}_n(t,x)\tilde{\sigma}_n(t',x')\rangle=n\int_{0}^{\lambda_m} d\lambda \lambda^{n-1}\mP_t(\lambda),
\eea 
and
\bea
&&\langle [\dot{\sigma}_n(t,x),\tilde{\sigma}_n(t',x')]\rangle=n\int_{0}^{\lambda_m} d\lambda \lambda^{n-1}\left(\mP_t(\lambda)-\mP_t^*(\lambda)\right)\nn \\
&&\phantom{\langle [\dot{\sigma}_n(t,x),\tilde{\sigma}_n(t',x')]\rangle}=2n\int_{0}^{\lambda_m} d\lambda \lambda^{n-1} \text{Im}\mP_t
\eea
The commutator is associated with the imaginary part of the function $\mP_t$. Using (\ref{imaginarypart_vacuum}) we have
\bea\label{commutator_vacuum}
\frac{\langle [\dot{\sigma}_n(t,x),\tilde{\sigma}_n(t',x')]\rangle}{\langle \sigma_n(t,x)\tilde{\sigma}_n(t',x')\rangle }=\frac{i\pi c}{6}\left(\frac{1}{n}-n\right)\left[ \delta(u-u')+\delta(v-v')\right].
\eea
If $n\sim 1$ we have
\bea
\langle [\dot{\sigma}_n(t,x),\tilde{\sigma}_n(t',x')]\rangle=\frac{i \pi c }{3}(1-n)\left[ \delta(u-u')+\delta(v-v')\right]+O(1-n)^2.
\eea

\section{Conclusion and discussion}\label{section_Conclusion}

In this paper, we further investigate the imaginary part of the timelike entanglement entropy (EE). In the previous work \cite{Guo:2024lrr}, it is  obtained the timelike EE through the analytical continuation of Euclidean correlation functions of twist operators. Here, we delve deeper into the details of this analytical continuation (\ref{Sn_separation}) and extend the relationship between timelike and spacelike EE to more general cases. This suggests that timelike and spacelike EE can be regarded as a unified framework.

The main purpose of this paper is to investigate the imaginary part of the timelike entanglement entropy (EE). Utilizing the relation (\ref{S_separation}), it was noted in \cite{Guo:2024lrr} that in certain special cases, the imaginary part is related to the commutators involving the twist operators and their first-order temporal derivatives. In this paper, we directly compute the commutators for the vacuum state, thermal state, and states dual to pure AdS$_3$. The results are consistent with the previous analysis, which is anticipated since all the aforementioned examples satisfy equation (\ref{Sn_separation}). It is also noteworthy that the imaginary part in these examples is universal.constant. 

To understand the properties of the imaginary part of timelike EE, we further explore the results for more general states using the OPE of twist operators in two-dimensional CFTs. Through calculations, we find that the commutator of the twist operator and its first-order derivative is a \textit{universal} constant that depends solely on the central charge. This clarifies why the examples considered in Section \ref{section_example} yield a universal imaginary part of the timelike EE when evaluated using equation (\ref{Im_0}).

However, in the most general cases, we expect that the imaginary part of the timelike EE is no longer constant. Instead, it should depend on the expectation values of the operators appearing in the OPE of the twist operator and the coordinates of the intervals. Consequently, the formula for the imaginary part of the timelike EE given in Eq.(\ref{Im_0}) is expected to receive corrections related to the OPE of the twist operators. In fact, we find it necessary to include commutators of the twist operators along with their higher-order temporal derivatives. In Section \ref{section_General}, we calculate these commutators that involve higher-order temporal derivatives. Utilizing the results from this section, we propose a modified formula for the imaginary part of the timelike EE that incorporates more general corrections from the OPE of the twist operator. We also verify the accuracy of this modified formula. Moreover, it is noteworthy that, in most general cases, the commutator of the twist operator and its first-order spatial derivative vanishes.

Extending the definition of timelike EE to higher dimensions is an intriguing question. There are certain subtleties regarding the definition of timelike EE for general subregions in higher dimensions. However, for the case of a strip, one can establish a well-defined timelike EE. Furthermore, it is possible to calculate the timelike EE using the holographic RT formula. This allows us to explore the imaginary part of the timelike EE for strips in a manner similar to the two-dimensional examples. In this paper, we calculate the commutator of the twist operator and its first-order derivative for the strip, as detailed in equation (\ref{commutator_hd}). Using this result, we derive a formula for the imaginary part of the timelike EE, presented in equation (\ref{Im_hd}). Our findings provide valuable insights into the origins of the imaginary part of the timelike EE in higher dimension.

Lastly, we demonstrate that the commutators of twist operators can be rephrased in terms of the spectra of the density matrix. The R\'enyi entropy and the entanglement spectra are interconnected through Laplace transforms, indicating that they contain the same information. This understanding can enhance our insights into various properties of timelike EE. If we consider timelike EE to be a special case of pseudo entropy, one might expect the existence of a non-Hermitian transition matrix. In general, its spectra should consist of complex numbers. An intriguing question arises regarding how to reconstruct the transition matrix using the timelike R\'enyi entropy.

There are many intriguing directions worth exploring in the near future. In this paper, we primarily focus on the field theory perspective. It would be fascinating to investigate how these results translate to the holographic setting. The commutator of the twist operator and its first-order temporal derivative is universal. This universality implies that the difference between the derivative of the Rényi entropy with respect to time $t$ and the derivative of its conjugate is always equal to $-8h_n\pi i\delta(x-x')$ for general state, that is
\begin{align}\label{Renyi_complex}
\p_t S_n(t,x;t',x')|_{t\to t_0,t'\to t_0}-\p_t S_n^*(t,x;t',x')|_{t\to t_0,t'\to t_0}=-8h_n\pi i\delta(x-x').
\end{align}
In the holographic framework, the holographic R\'enyi entropy is connected to the bulk action through the replica method \cite{Lewkowycz:2013nqa}. In \cite{Colin-Ellerin:2020mva,Colin-Ellerin:2021jev}, the authors developed a real-time replica method to evaluate holographic R\'enyi entropy. The relation given in Eq.(\ref{Renyi_complex}) appears to necessitate a complex bulk action. It would be intriguing to explore how to derive the universal relation (\ref{Renyi_complex}) from the bulk perspective.

Furthermore, the universal commutator of the twist operator and its first-order temporal derivative bears a striking resemblance to the commutators of field operators and their conjugate momenta in the canonical quantization of field theory. It would be a compelling direction to investigate whether this universal commutator relation can be used to quantize both field and bulk theories.

In this paper, we focus solely on the imaginary part of the timelike EE. An intriguing question arises regarding the formulation of a complete connection between timelike and spacelike EE, which would serve as a correction to the formula given in Eq.(\ref{S_separation}). We plan to explore this question in the near future. For the higher-dimensional case, we have identified an extension of the imaginary part of the timelike EE for the strip example. It would be interesting to extend these results to more general scenarios involving more complex subregions.


~\\
~\\
{\bf Acknowledgements}
We would like to thank Zheng-Quan Cui, Song He, Rong-Xin Miao, Run-Qiu Yang and Jiaju Zhang for the useful discussions.
WZG is supposed by the National Natural Science Foundation of China under Grant No.12005070.

\appendix

\section{Correlators of twist operators involving time derivative}\label{app1}
By defintion the operator $\dot{\sigma}_n$ is given by
\bea
\dot{\sigma}_n= i[H^{(n)},\sigma_n],
\eea
where $H^{(n)}$ is the Hamiltonian of CFT$_n$. The Hamiltonian can be written as 
\bea\label{Hamiltonian}
H^{(n)}=\int dx_0 T^{(n)}_{00}(0,x_0),
\eea
where $T_{00}$ is the energy density.
The commutator $\langle \Psi| [\dot{\sigma}_n(t,x),\tilde{\sigma}_n(t',x')]|\Psi\rangle$ for general state can be expressed as
\bea
&&i\langle \Psi| [[H^{(n)},\sigma_n(t,x)],\tilde{\sigma}_n(t',x')]|\Psi\rangle\nn \\
&&=i\langle \Psi| H^{(n)}\sigma_n(t,x)\tilde{\sigma}_n(t',x')|\Psi\rangle-i \langle \Psi|\sigma_n(t,x) H^{(n)}\tilde{\sigma}_n(t',x')|\Psi\rangle\nn \\
&&-i \langle \Psi| \tilde{\sigma}_n(t',x') H^{(n)}\sigma_n(t,x)|\Psi\rangle+i\langle \Psi| \tilde{\sigma}_n(t',x') \sigma_n(t,x)H^{(n)}|\Psi\rangle.
\eea
Using (\ref{Hamiltonian}) the above results can be computed by correlators involving the energy density operator $T_{00}$. To achieve this we can firstly consider the Eunclidean correlators then obtain the Lorentzian ones by analytical continuation.

For general states it is hard to directly evaluate the correlator (\ref{twist}). We can turn to using the Ward identity. We would like to consider the correlator involving the stress energy tensor of $T^{(n)}$ and $\bar T^{(n)}$ of the theory CFT$_n$,
\bea\label{Wardgeneral}
\langle \Psi|T^{(n)}(z)\sigma_n(\tau,x)\tilde{\sigma}_n(\tau',x')|\Psi\rangle,\quad \langle \Psi|\bar T^{(n)}(z)\sigma_n(\tau,x)\tilde{\sigma}_n(\tau',x')|\Psi\rangle,
\eea 
where $T^{(n)}=\sum_i T_i$, $T_i$ is the stress energy tensor for the theory of the $i$-th copy. For $|\psi\rangle=|0\rangle$ one could evaluate the above correlator by Ward identity. 

Assume the position of $T^{(n)}$ and $\bar T^{(n)}$ is near the point $(\tau,x)$. By using Ward identity the leading contributions of the correlator (\ref{Wardgeneral}) are given by
\bea
&&\langle \Psi|T^{(n)}(z)\sigma_n(\tau,x)\tilde{\sigma}_n(\tau',x')|\Psi\rangle
\nn\\
&&=\frac{h_n}{(z-w)^2}\langle \Psi|\sigma_n(\tau,x)\tilde{\sigma}_n(\tau',x')|\Psi\rangle+\frac{1}{z-w}\p_w\langle \Psi|\sigma_n(\tau,x)\tilde{\sigma}_n(\tau',x')|\Psi\rangle+...\;,\nn\\
&&\langle \Psi|\bar T^{(n)}(\bar z)\sigma_n(\tau,x)\tilde{\sigma}_n(\tau',x')|\Psi\rangle
\nn\\
&&=\frac{h_n}{(\bar z-\bar w)^2}\langle \Psi|\sigma_n(\tau,x)\tilde{\sigma}_n(\tau',x')|\Psi\rangle+\frac{1}{\bar z-\bar w}\p_w\langle \Psi|\sigma_n(\tau,x)\tilde{\sigma}_n(\tau',x')|\Psi\rangle+...\;,\nn
\eea
where $z:=\tau_0+i x_0$, $\bar z:= \tau_0-i x_0$ and $w:=\tau+ix$, $\bar w:=\tau-i x$. If the stress-energy tensor is located in proximity to $(t', x')$, a comparable outcome would be anticipated. One could obtain the Lorentzian correlators by analytical continuation $\tau\to it+\epsilon$. 

\section{Details for the different separations}\label{Details_separations}
For the separation in Fig.\ref{separation} (a), we know $-u<-u'<v'<v$, so we have
\begin{align}
&\frac{1}{4}\int_{-u'}^{v'}\text{d}y \p_{t'}S_n(t_0,-u;t_0,y)+\frac{1}{4}\int_{-u'}^{v'}\text{d}y \p_{t'}S_n(t_0,v;t_0,y)\nn\\
=&-\frac{c}{48}(1+\frac{1}{n})\{\int_{-u'}^{v'}\text{d}y [\frac{1}{-u-y-i\epsilon}-\frac{1}{-u-y+i\epsilon}]+\int_{-u'}^{v'}\text{d}y [\frac{1}{v-y-i\epsilon}-\frac{1}{v-y+i\epsilon}]\}\nn\\
=&-\frac{c}{48}(1+\frac{1}{n})\{\int_{-u'}^{v'}\text{d}y [2\pi i\delta(-u-y)]+\int_{-u'}^{v'}\text{d}y [2\pi i\delta(v-y)]\}\nn\\
=&0,
\end{align}
and
\begin{align}
&\frac{1}{4}\int_{-u}^{v}\text{d}y \p_{t}S_n(t_0,y;t_0,-u')+\frac{1}{4}\int_{-u}^{v}\text{d}y \p_{t}S_n(t_0,y;t_0,v')\nn\\
=&\frac{c}{48}(1+\frac{1}{n})\{\int_{-u}^{v}\text{d}y [\frac{1}{y+u'-i\epsilon}-\frac{1}{y+u'+i\epsilon}]+\int_{-u}^{v}\text{d}y [\frac{1}{y-v'-i\epsilon}-\frac{1}{y-v'+i\epsilon}]\}\nn\\
=&\frac{c}{48}(1+\frac{1}{n})\{\int_{-u}^{v}\text{d}y [2\pi i\delta(y+u')]+\int_{-u}^{v}\text{d}y [2\pi i\delta(y-v')]\}\nn\\
=&\frac{i\pi c}{12}(1+\frac{1}{n}).
\end{align}
For the separation in Fig.\ref{separation} (b), we know $-u'<-u<v<v'$, so we have
\begin{align}
&\frac{1}{4}\int_{-u'}^{v'}\text{d}y \p_{t'}S_n(t_0,-u;t_0,y)+\frac{1}{4}\int_{-u'}^{v'}\text{d}y \p_{t'}S_n(t_0,v;t_0,y)\nn\\
=&\frac{c}{48}(1+\frac{1}{n})\{\int_{-u'}^{v'}\text{d}y [2\pi i\delta(-u-y)]+\int_{-u'}^{v'}\text{d}y [2\pi i\delta(v-y)]\}\nn\\
=&-\frac{i\pi c}{12}(1+\frac{1}{n}),
\end{align}
and
\begin{align}
&\frac{1}{4}\int_{-u}^{v}\text{d}y \p_{t}S_n(t_0,y;t_0,-u')+\frac{1}{4}\int_{-u}^{v}\text{d}y \p_{t}S_n(t_0,y;t_0,v')\nn\\
=&-\frac{c}{48}(1+\frac{1}{n})\{\int_{-u}^{v}\text{d}y [2\pi i\delta(y+u')]+\int_{-u}^{v}\text{d}y [2\pi i\delta(y-v')]\}\nn\\
=&0.
\end{align}
For the separation in Fig.\ref{separation} (c), we know $-u<v<-u'<v'$, so we have
\begin{align}
&\frac{1}{4}\int_{-u'}^{v'}\text{d}y \p_{t'}S_n(t_0,-u;t_0,y)+\frac{1}{4}\int_{-u'}^{v'}\text{d}y \p_{t'}S_n(t_0,v;t_0,y)\nn\\
=&-\frac{c}{48}(1+\frac{1}{n})\{\int_{-u'}^{v'}\text{d}y [2\pi i\delta(-u-y)]+\int_{-u'}^{v'}\text{d}y [2\pi i\delta(v-y)]\}\nn\\
=&0,
\end{align}
and
\begin{align}
&\frac{1}{4}\int_{-u}^{v}\text{d}y \p_{t}S_n(t_0,y;t_0,-u')+\frac{1}{4}\int_{-u}^{v}\text{d}y \p_{t}S_n(t_0,y;t_0,v')\nn\\
=&\frac{c}{48}(1+\frac{1}{n})\{\int_{-u}^{v}\text{d}y [2\pi i\delta(y+u')]+\int_{-u}^{v}\text{d}y [2\pi i\delta(y-v')]\}\nn\\
=&0.
\end{align}
For the separation in Fig.\ref{separation} (d), we know $-u<-u'<v<v'$, so we have
\begin{align}
&\frac{1}{4}\int_{-u'}^{v'}\text{d}y \p_{t'}S_n(t_0,-u;t_0,y)+\frac{1}{4}\int_{-u'}^{v'}\text{d}y \p_{t'}S_n(t_0,v;t_0,y)\nn\\
=&-\frac{c}{48}(1+\frac{1}{n})\{\int_{-u'}^{v'}\text{d}y [2\pi i\delta(-u-y)]+\int_{-u'}^{v'}\text{d}y [2\pi i\delta(v-y)]\}\nn\\
=&-\frac{i\pi c}{24}(1+\frac{1}{n}),
\end{align}
and
\begin{align}
&\frac{1}{4}\int_{-u}^{v}\text{d}y \p_{t}S_n(t_0,y;t_0,-u')+\frac{1}{4}\int_{-u}^{v}\text{d}y \p_{t}S_n(t_0,y;t_0,v')\nn\\
=&\frac{c}{48}(1+\frac{1}{n})\{\int_{-u}^{v}\text{d}y [2\pi i\delta(y+u')]+\int_{-u}^{v}\text{d}y [2\pi i\delta(y-v')]\}\nn\\
=&\frac{i\pi c}{24}(1+\frac{1}{n}).
\end{align}
\section{Details for the commutator of Thermal state}\label{Details_Thermal}
Since we have
\begin{align}
\frac{\langle \dot{\sigma}_n(t,x)\tilde{\sigma}_n(t',x')\rangle_{\beta}}{\langle \sigma_n(t,x)\tilde{\sigma}_n(t',x')\rangle_{\beta}}=\frac{-2h_n\cosh{(\frac{\pi(u-u'-i\epsilon)}{\beta}})}{\frac{\beta}{\pi}\sinh{(\frac{\pi(u-u'-i\epsilon)}{\beta}})}+\frac{-2\bar h_n\cosh{(\frac{\pi(v-v'-i\epsilon)}{\beta}})}{\frac{\beta}{\pi}\sinh{(\frac{\pi(v-v'-i\epsilon)}{\beta}})}
\end{align}
The two terms are similar, so we might as well analyze the term involving $u$ first, we have
\begin{align}
&-2h_n[\frac{\cosh{(\frac{\pi(u-u'-i\epsilon)}{\beta}})}{\frac{\beta}{\pi}\sinh{(\frac{\pi(u-u'-i\epsilon)}{\beta}})}-\frac{\cosh{(\frac{\pi(u-u'+i\epsilon)}{\beta}})}{\frac{\beta}{\pi}\sinh{(\frac{\pi(u-u'+i\epsilon)}{\beta}})}]\nn\\
=&-2h_n[\frac{\cosh{(\frac{\pi(u-u'-i\epsilon)}{\beta}})}{\frac{\beta}{\pi}\sinh{(\frac{\pi(u-u'-i\epsilon)}{\beta}})}-\frac{\cosh{(\frac{\pi(u-u'+i\epsilon)}{\beta}})}{\frac{\beta}{\pi}\sinh{(\frac{\pi(u-u'+i\epsilon)}{\beta}})}]\nn\\
=&-2h_n\frac{\sinh{(\frac{\pi}{\beta}2i\epsilon})}{\frac{\beta}{\pi}\sinh{(\frac{\pi(u-u'-i\epsilon)}{\beta}})\sinh{(\frac{\pi(u-u'+i\epsilon)}{\beta}})}.
\end{align}
When $u-u'\neq0$, the denominator is a non-zero constant, and the numerator tends to $0$. Therefore, both the correlator and the commutator are 0. And when $u-u'\to 0$, we have
\begin{align}
&-2h_n\frac{\sinh{(\frac{\pi}{\beta}2i\epsilon})}{\frac{\beta}{\pi}\sinh{(\frac{\pi(u-u'-i\epsilon)}{\beta}})\sinh{(\frac{\pi(u-u'+i\epsilon)}{\beta}})}\nn\\
=&-2h_n\frac{\frac{\pi}{\beta}2i\epsilon}{\frac{\beta}{\pi}\frac{\pi(u-u'-i\epsilon)}{\beta}\frac{\pi(u-u'+i\epsilon)}{\beta}}\nn\\
=&-2h_n[\frac{1}{u-u'-i\epsilon}-\frac{1}{u-u'+i\epsilon}]\nn\\
=&-4h_n\pi i\delta(u-u')
\end{align}
Combining the terms of $u$ and $v$, the commutator is
\begin{align}
\frac{\langle [\dot{\sigma}_n(t,x),\tilde{\sigma}_n(t',x')]\rangle_{\beta}}{\langle \sigma_n(t,x)\tilde{\sigma}_n(t',x')\rangle_{\beta}}=-4h_n[i\pi\delta(u-u')+i\pi\delta(v-v')].
\end{align}

\end{document}